\documentclass[10pt,journal,compsoc]{IEEEtran}
\usepackage{times}
\usepackage{epsfig}
\usepackage{graphicx}
\usepackage{amsmath}
\usepackage{amssymb}
\usepackage{algorithm}
\usepackage{algorithmic}
\usepackage[pagebackref=true,breaklinks=true,letterpaper=true,colorlinks,bookmarks=false]{hyperref}

\newtheorem{claim}{Claim}

\ifCLASSOPTIONcompsoc
  \usepackage[nocompress]{cite}
\else
  \usepackage{cite}
\fi
\ifCLASSINFOpdf
\else
\fi
\begin{document}
%
\title{Optimal HDR and Depth from Dual Cameras}
%
%
%
%

\author{Pradyumna~Chari,
        Anil~Kumar~Vadathya,
        and~Kaushik~Mitra
        
        \small{E-mail: pradyumnac@ucla.edu, av57@rice.edu, kmitra@ee.iitm.ac.in}
}

\IEEEtitleabstractindextext{%
\begin{abstract}
Dual camera systems have assisted in the proliferation of various applications, such as optical zoom, low-light imaging and High Dynamic Range (HDR) imaging. In this work, we explore  an optimal method for capturing the scene HDR and disparity map using dual camera setups. 
Hasinoff et al. \cite{Hasinoff2010NoiseoptimalCF} have developed a noise optimal framework for HDR capture from a single camera. We generalize this to the dual camera set-up for estimating both HDR and disparity map. It may seem that dual camera systems can capture HDR in a shorter time. However, disparity estimation is a necessary step, which requires overlap among the images captured by the two cameras. This might lead to an increase in the capture time.
To address this conflicting requirement, we propose a novel framework to find the optimal exposure and ISO sequence by \textit{minimizing the capture time} under the constraints of an \textit{upper bound on the disparity error} and a \textit{lower bound on the per-exposure SNR}. We show that the resulting optimization problem is non-convex in general and propose an appropriate initialization technique. To obtain the HDR and disparity map from the optimal capture sequence, we propose a pipeline which alternates between estimating the camera ICRFs and the scene disparity map. We demonstrate that our optimal capture sequence leads to better results than other possible capture sequences. Our results are also close to those obtained by capturing the full stereo stack spanning the entire dynamic range. Finally, we present for the first time a stereo HDR dataset consisting of dense ISO and exposure stack captured from a smartphone dual camera. The dataset consists of $6$ scenes, with an average of $142$ exposure-ISO image sequence per scene. The Supplementary material for this work may be found \href{https://tinyurl.com/stereoHDRsupp}{here}.
\end{abstract}
}
\maketitle

\IEEEdisplaynontitleabstractindextext

\IEEEpeerreviewmaketitle

\IEEEraisesectionheading{\section{Introduction}\label{sec:intro}}
\IEEEPARstart{I}{n} recent years dual cameras have gained ubiquity among consumer devices, especially, in smartphones. Such imaging setups allow for new capabilities towards scene capture. Two such capabilities are of particular importance: (a) simultaneous image capture, and (b) view diversity. These have led to the proliferation of various applications of dual cameras such as optical zoom photography, low-light imaging and scene HDR capture. In this work, we explore dual cameras for capturing both the scene HDR and disparity map. This will enable downstream applications such as refocusing and view synthesis of HDR images. 


HDR imaging is a vastly studied problem in computational imaging, where the typical approach is to capture multiple images of the scene with varying exposure time \cite{Debevec97recoveringhigh,robertson2003estimation, manders2004camera, barakat2008minimal}. Each of these images captures a portion of the dynamic range, which are then fused to obtain the HDR image. Traditionally, exposure stack for HDR is captured with a varying exposure factor of $2$ and a nominal constant value of ISO ($100$ or $200$) to suppress noise. However, Hasinoff et al.~\cite{Hasinoff2010NoiseoptimalCF} have shown that this is non-optimal in terms of capture time, and have proposed varying the ISO along with the exposure time based on their noise analysis. Along similar lines, we propose a framework for finding the optimal exposure and ISO sequence for dual camera setups. This differs from \cite{Hasinoff2010NoiseoptimalCF} in the aspect that we also need to estimate the scene disparity map along with the HDR image.  

If we were only interested in capturing the scene HDR, then the simultaneous use of dual cameras should reduce the total capture time. 
However, the need for disparity estimation requires that the stereo pair should capture overlapping radiance values. Thus, an arbitrary assignment of exposure times and ISO control among dual cameras does not lead to a successful HDR and depth estimation. We therefore generalize our overall objective to include both disparity and HDR estimation. Considering these conflicting requirements of HDR recovery and depth estimation, we propose a framework to find an optimal capture sequence under the following constraints: the enitre radiance range of interest should be covered, capture time should be \textit{minimized} , disparity error should be \textit{minimized} and per-capture SNR should be greater than a minimum threshold.

\begin{figure*}[t]
    \centering
    \includegraphics[width=\textwidth]{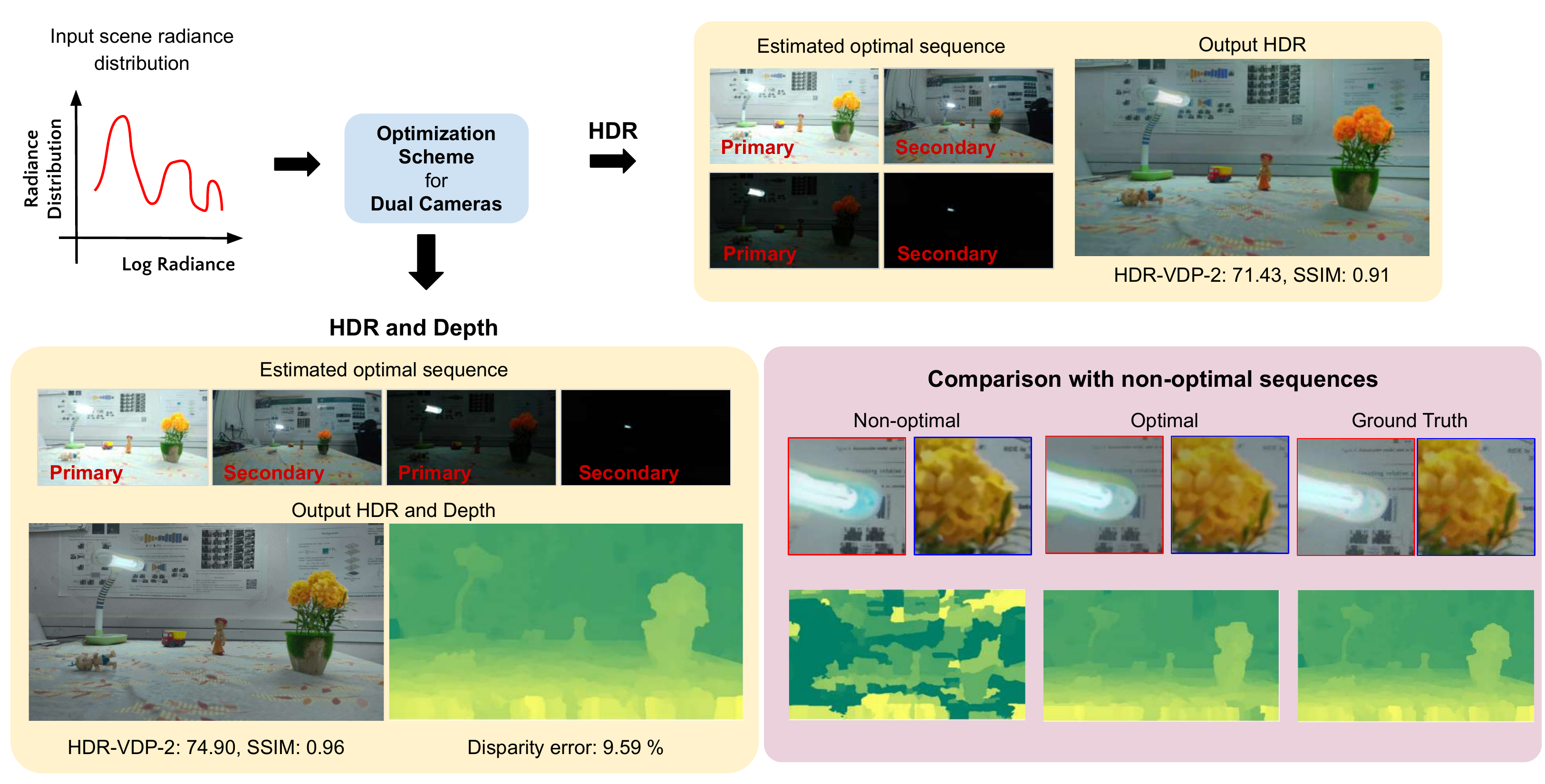}
    \caption{Given scene radiance distribution, our framework computes the optimal exposure and ISO sequence for dual camera setups. Our framework can handle both the cases of: i) estimating just HDR and ii) jointly estimating HDR and disparity map. 
    Our reconstructions are similar to the ground-truth reconstructions obtained from the full exposure stack. 
    }
    \label{fig:main}
\end{figure*}

Our algorithm uses the scene radiance distribution as an input to ensure that the full radiance range of the scene is covered.
Further, the radiance distribution is also used to estimate the fraction of pixels with erroneous disparity (hereafter referred to as disparity error). The pixels whose radiance values are not captured without saturation by both the cameras are essentially the pixels which give rise to erroneous disparity values. Once we have the optimal input capture sequence, we iteratively estimate the inverse camera response functions (ICRFs) for the two cameras, and the disparity map. These are then used to obtain the optimally fused HDR image, as well as the depth map. We show that the HDR and depth representation obtained from the optimal sequence is qualitatively and quantitatively better than those obtained from most other naively chosen capture sequences. 
As part of our work we release a stereo HDR dataset of $6$ scenes, with each scene consisting of the full exposure and ISO stack captured using the dual camera of a LG G5 smartphone. To summarize, the major contributions of this work are as follows:

\begin{itemize}
    \item We propose a framework for recovering HDR and depth from dual cameras. We analyze the conflicting requirements for these tasks and present a generalized optimization scheme to minimize the capture time and disparity error. 
    \item We show that the optimization problem at hand (specifically the disparity error constraint) is non-convex in general. Therefore, we additionally propose an appropriate initialization scheme for our algorithm. 
    \item We propose a pipeline to jointly estimate Inverse Camera Response Functions (ICRFs), scene disparity map and the HDR image from the input image sequence.
    \item We demonstrate the optimality of our approach over other possible capture schemes. 
    \item We present a Stereo HDR dataset consisting of $6$ scenes captured using the LG G5 cellphone dual cameras. 
    We will release this dataset upon publication of the paper.
\end{itemize}

\section{Previous Work}\label{sec:prev_work}

\textbf{HDR imaging:} High dynamic range imaging through fusion of multiple Low Dynamic Range (LDR) images is a well-researched topic in the domain of computational photography. Several previous works have looked into various aspects contained within HDR imaging, namely identifying required exposures \cite{Pourreza}, conversion of images to radiance space by identifying and inverting the camera response function \cite{Debevec97recoveringhigh, grossberg2003high}, and fusion of scene radiance information from multiple LDR images to obtain a single HDR image \cite{Robertson99dynamicrange, madden1993extended, yamada1994wide, moriwaki1994adaptive}. Additionally, deghosting or the removal of motion induced artifacts is also well explored in works such as \cite{HuDeghost, gallo2009artifact, heo2010ghost}. More recently, several deep learning based HDR frameworks, both for multi-image HDR and single-image HDR have been successful in obtaining state of the art results \cite{EKDMU17,LearningHDR, wu2018deep, lee2018deep, cai2018learning}.

Kalantari et al. \cite{LearningHDR} proposed a deep learning based solution for HDR reconstruction for dynamic scenes. They use the popular approach of exposure bracketing to capture an input sequence of three LDR images with motion. The motion could be due to camera motion or because of dynamic scenes. They use optical flow to align input LDR images to the reference medium exposure image. The aligned images are used as input for learning based HDR fusion which is then tone mapped to give the final HDR image. 

\noindent\textbf{Finding optimal capture sequence for single camera HDR:}
The idea of optimally selecting exposures and ISOs for imaging was previously proposed in \cite{Hasinoff2010NoiseoptimalCF}. The optimality, in their case, is established based on the notion of maximizing SNR and/or minimizing capture time. 
Their results show performance improvement (either in terms of capture time or final HDR image noise performance, depending on the objective function) over other naive single camera capture sequences. Additionally, the work provides experimental backing for the possible SNR advantages to be gained from ISO control.

\noindent\textbf{HDR from Stereo:}
Using stereo camera setups for HDR imaging has been explored in past works \cite{tomaszewska2007image, troccoli2006multi, sun2010hdr}. Early works of Tomaszewska et al. \cite{tomaszewska2007image} look at registration aspects to address misalignment in the handheld images taken at different exposures.  Troccolli et al. \cite{troccoli2006multi} propose an exposure invariant matching method for HDR and depth recovery from multi-view multi-exposure setting. Lin et al. \cite{Lin:2009:HDR:1819298.1819909} propose SIFT based feature matching for registration of stereo pairs with different exposures. Batz et al. \cite{Batz:2014:HDR:2583125.2583269} extend the formulation to account for stereo HDR for videos. Park et al. \cite{ParkHDR} look at improving the pipeline for Stereo HDR over the previously works by incorporating hole-filling algorithms for occluded regions. Hafner et al. \cite{hafner2014simultaneous} propose an approach for simultaneous HDR and optical flow to align the input LDR images for motion compensation. Note that all of these approaches ignore the input acquisition part and start with a generic stereo exposure stack irrespective of the scene radiance distribution. In this work we propose an optimization framework for finding the optimal exposure and ISO sequence for HDR recovery and depth reconstruction from dual camera. 

\noindent\textbf{Depth from Stereo:} Depth estimation using stereo cameras is a widely studied topic \cite{hirschmuller2005accurate, liang2018learning, liang2019stereo}. The Middlebury Stereo Evaluation framework \cite{Hirschmuller07eval} provides a comprehensive performance analysis for various disparity estimation algorithms. In this work, we do not focus on finding the optimal disparity estimation algorithm for our purpose. Instead our goal is to make our framework flexible with respect to disparity estimation algorithms. However, in our experiments, we use the disparity estimation algorithm proposed by Mozerov et al. \cite{MozerovDisp}.

\section{Optimal Framework for HDR and Depth}
\label{OptFrame}

\subsection{Log Radiance Intervals for HDR}
\label{subsec:CRF}
An imaging sensor works by mapping the scene radiance to a value between $0$ to $2^{n}-1$, where $n$ is the bit-depth for the camera. This is given by:

\begin{equation}
    d = f(\frac{\phi t}{g}),
    \label{eq:I_sens}
\end{equation}

\noindent when not considering additive noise. Here $d$ is the image pixel value, $f(\cdot)$ is the monotonic camera response function (CRF), $\phi$ is the radiance of the scene point, $t$ is the exposure duration and $g$ is the gain of the camera, which is inversely related to ISO. Let $x_l$ and $x_u$ be the lower and upper range of $\frac{\phi t}{g}$ for which the pixel values are neither too noisy nor over saturated, respectively. Then, the useful range of $\frac{\phi t}{g}$ is given by
\begin{equation}
    x_l < \frac{\phi t}{g} < x_u.
    \label{eq:Useful_range}
\end{equation}

Note that $x_l$ is a function of ISO, i.e. $x_l$ increases with ISO. This aspect is further discussed in Sections~\ref{subsec:iso} and \ref{subsec:init}. An image captured with a particular exposure time $t$ and gain $g$ can be interpreted as capturing a particular range of radiance values on the log radiance scale, subsequently referred to as \textit{log radiance interval} in this work. 
This is obtained by applying logarithm to Equation \ref{eq:Useful_range}:
\begin{equation}
    log(x_l)-log(t)+log(g)<log(\phi)<log(x_u)-log(t)+log(g).
\end{equation}
The optimal HDR capture problem reduces to identifying a set of exposures and ISOs, so as to span the dynamic range of interest. For our purpose, we look to estimate the log inverse camera response function (ICRF), since we wish to go from pixel values to radiance estimates. For brevity, this is referred to as ICRF in the rest of the paper, using the notation $e(\cdot)=log f^{-1}(\cdot)$, with appropriate subscripts/superscripts as required.

\begin{figure}[t]
\begin{center}
\includegraphics[width=7cm]{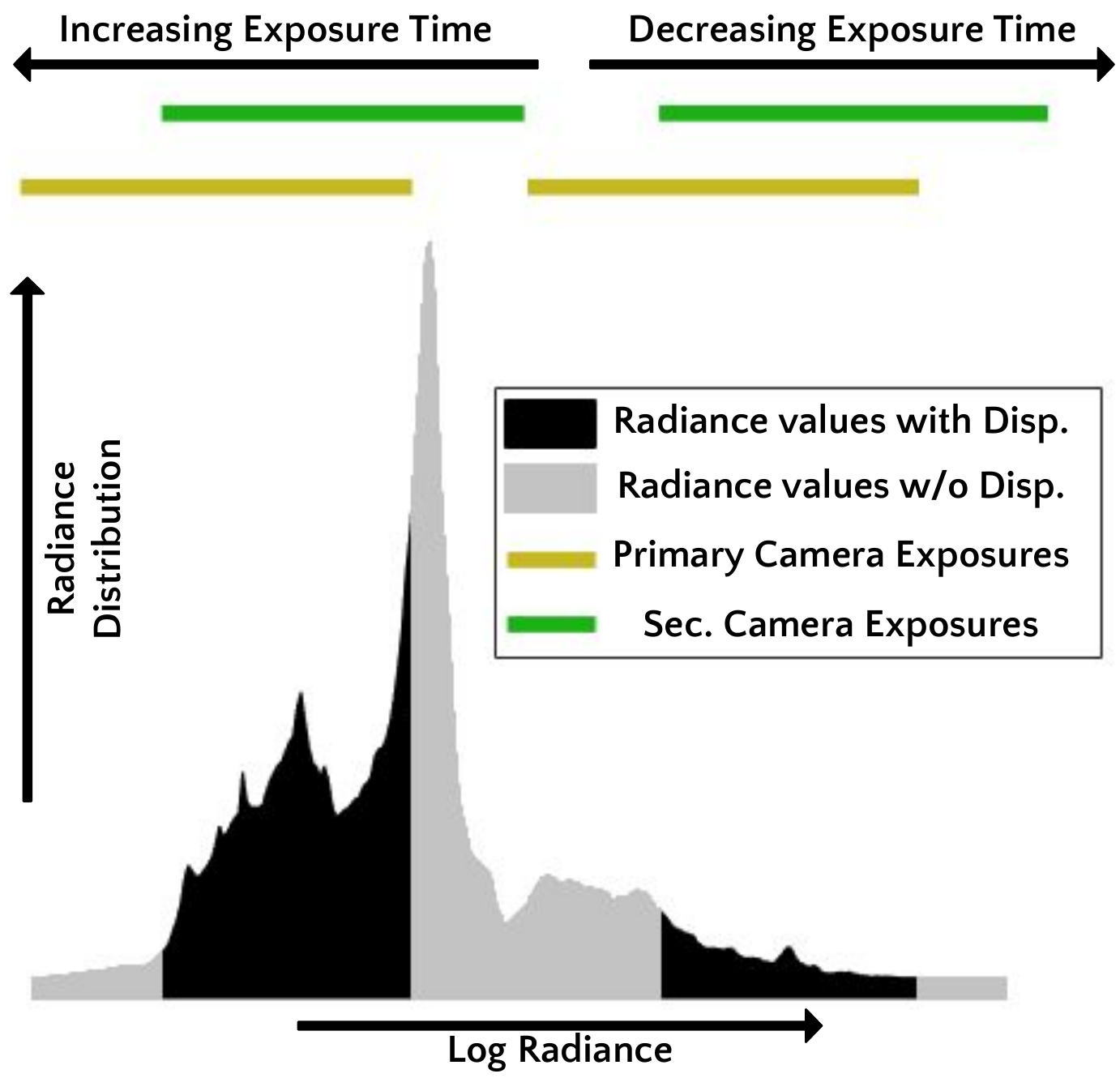}
\end{center}
\caption{Scene Description and Camera Performance Analysis Using Radiance Distribution Maps. The yellow and green brackets represent the log radiance interval captured by the primary and secondary cameras respectively at a given exposure and ISO setting. Reducing the exposure moves the bracket right, while increasing the exposure moves the bracket left. HDR reconstruction can be viewed as a problem of optimal placement of these brackets, while disparity estimation can be viewed as a problem of maximizing overlap between these brackets.}
\label{fig:radMap}
\end{figure}





\begin{figure*}[t]
\centering
\includegraphics[width=17cm]{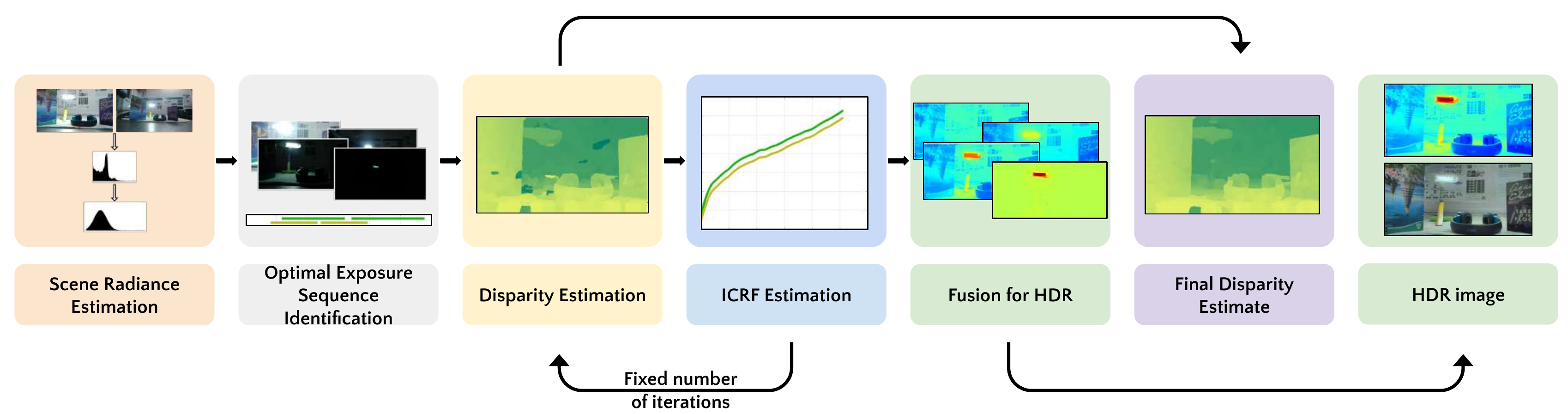}
\caption{Description of the proposed pipeline. The steps include: (i) Estimating the scene radiance distribution, (ii) finding the images to be captured (in terms of exposure times and ISOs) using our optimization framework, (iii) iteratively estimating disparity (using \cite{MozerovDisp}) and ICRFs (using our method) for a fixed number of iterations, depending on required accuracy and robustness of disparity estimation algorithm, and (iv) fusion of the radiance information using our fusion paradigm.}
\label{fig:framework}
\end{figure*}

\subsection{Optimal HDR and Depth Estimation Criteria}
\label{subsec:criteria}
Our optimization framework is mainly based on the knowledge of scene radiance distribution. Any scene of interest can be represented on the basis of its radiance distribution, which is a histogram of the various radiance values in the scene. In order to capture the HDR image, the entire dynamic range of interest must be spanned, such that each radiance is captured, without saturation, by at least one of the dual cameras. Additionally, since we are dealing with a stereo camera setup, we are also interested in estimating the disparity map. This is required both for depth estimation as well as for registering images for HDR reconstruction. 
For successfully estimating the disparity value of a point in the scene, it must be captured without saturation in both the cameras.

To better understand the constraints the exposure sequence on dual cameras should satisfy refer to Figure \ref{fig:radMap}. It shows an example scene radiance distribution, along with the radiance range covered by the various captured images from the dual cameras. The radiance values where the exposures do not overlap between the primary camera (camera whose viewpoint is of final interest to us) and the secondary camera contribute to the disparity error, since stereo matching will not work when the intensity values are quite different. The exposures should therefore be distributed among the stereo cameras such that the range of radiance values captured on left and right cameras show reasonable overlap. Hence, one of our objectives is to keep the fraction of pixels with disparity error below a threshold. Note that there could be additional error in disparity estimation which could be because of lack of texture in the scene or even could depend on the specific algorithm that we use for disparity estimation. To accommodate for such errors, we can keep our threshold a bit lower than what we finally desire. 

Overall, we identify the following criteria for optimal HDR recovery and disparity estimation:
\begin{itemize}
    \item The capture time for the images should be minimum. We assume that simultaneous capture is possible from both the cameras. 
    \item Every radiance value within the dynamic range of interest must be captured, without saturation and with sufficiently low noise, in at least one image. This is to satisfy the coverage criterion.
    \item The fraction of scene pixels that are not properly captured by any of the cameras should be less than a threshold value. This is needed for accurate disparity estimation as well as for HDR recovery.
    \item The worst case SNR for each image should be greater than a specified threshold. 
\end{itemize}

We now set up the optimization framework for a stereo camera capture setup. We first establish the optimization framework without ISO control, for ease of understanding. ISO is then included in order to complete the discussion.

\subsection{Optimization Framework (without ISO)}
\label{subsec:optNoISO}
Let the various image exposures be represented as $t^j_i$, for the $i^{th}$ image in the $j^{th}$ camera. Assume that $m$ images are captured from the primary camera, and $n$ images captured using the secondary camera. Then, the overall capture time is:
\begin{equation}
    t_{cap}=max(\Sigma_{i=1}^m t^1_i,\Sigma_{i=1}^n t^2_i).
\end{equation}
This is because both the cameras may be operated simultaneously. 

Let $K^j_i$ represent the log radiance interval captured by the $i^{th}$ image from the $j^{th}$ camera. Additionally, let $R$ represent the radiance range of interest, which we wish to capture using our imaging setup. Then, the various log radiance intervals must satisfy
\begin{equation}
    (\cup_{i=1}^mK^1_i)\cup(\cup_{i=1}^nK^2_i)\supseteq R.
\end{equation}

Let $h(\cdot)$ represent the probability distribution function corresponding to the log radiance histogram for the scene, and let $\gamma_{err} \in 0\leq \gamma_{err}\leq1$ be the allowed error in disparity. Then, the disparity error criterion reduces to:
\begin{equation}
    1-\int_{O}h(x)dx\leq \gamma_{err},\hspace{0.2cm}    O = (\cup_{i=1}^mK^1_i)\cap(\cup_{i=1}^nK^2_i).
    \label{eq:disp_err_const}
\end{equation}

Based on the above description of the various factors involved, we now set up the optimization framework as follows:
\begin{equation}
\label{eq:opt}
    \begin{split}
        & \text{Minimize} \hspace{0.2cm} t_{cap}\\
        & \text{Subject to}\\
        & \hspace{0.2cm} (\cup_{i=1}^mK^1_i)\cup(\cup_{i=1}^nK^2_i)\supseteq R\\
        & \hspace{0.2cm} 1-\int_{O}h(x)dx\leq \gamma_{err}, \hspace{0.2cm}    O = (\cup_{i=1}^mK^1_i)\cap(\cup_{i=1}^nK^2_i).\\
    \end{split}
\end{equation}
The set of exposures for each camera that arise out of this optimization specify the optimal exposure capture sequence.

\subsection{Optimization Framework (with ISO)}
\label{subsec:iso}

Increasing the ISO allows for capturing a certain radiance range with a lower exposure time, which reduces the capture time. However, the noise characteristics of the image change with changing ISO. Based on the assumption of linear ICRF, Hasinoff et al. \cite{Hasinoff2010NoiseoptimalCF} have derived the Signal to Noise Ratio (SNR) for an unsaturated pixel as:
\begin{equation}
    SNR^j(\phi,t^{j}_{i},g^{j}_{i}) = \frac{\phi^2 {t^{j}_{i}}^2}{\phi {t^{j}_{i}}+\sigma_{r}^2+\sigma_{q}^2{g^{j}_{i}}^2},
    \label{eq:SNR}
\end{equation}

\noindent where $\phi$ is the scene point radiance, $j$ is the camera index, $t_{i}^{j}, g_{i}^{j}$ is the exposure duration and the sensor gain of the $i^{th}$ image respectively for the $j^{th}$ camera, $\sigma_r$ is the read noise and $\sigma_q$ is the quantization noise. Note that the ISO setting and the sensor gain are related as $ISO=\frac{K}{g}$, where $K$ is a camera-dependent constant. Additionally, for a given log radiance interval to be captured, the sensor gain and exposure time must satisfy $\frac{t}{g}=constant$. 

The aspect of ISO control can now be included in the optimization framework to further allow for freedom to reduce capture time. We propose a lower threshold on the worst-case (minimum) SNR $\eta$ (similar to \cite{Hasinoff2010NoiseoptimalCF}) for each of the captured images. Using this, the optimized framework can be rewritten as:
\begin{equation}
\label{eqn:optISO}
    \begin{split}
        & \text{Minimize} \hspace{0.2cm} t_{cap}\\
        & \text{Subject to}\\
        & \hspace{0.1cm} (\cup_{i=1}^mK^1_i)\cup(\cup_{i=1}^nK^2_i)\supseteq R\\
        & \hspace{0.1cm} 1-\int_{O}h(x)dx\leq \gamma_{err}, \hspace{0.2cm}    O = (\cup_{i=1}^mK^1_i)\cap(\cup_{i=1}^nK^2_i)\\
        & \hspace{0.1cm}
        \displaystyle{\min_{\phi \in K^j_i}  \hspace{0.1cm} SNR^{j}(\phi,t^{j}_{i},g^{j}_{i})}\geq \eta, \hspace{0.1cm} i\in \{1,...,m \}, \hspace{0.1cm}j\in \{1,2 \}.
    \end{split}
\end{equation}
Note that the presence of a tunable disparity error threshold allows our method to uniquely tackle both the only HDR as well as the joint HDR and disparity map recovery tasks. This is achieved by choosing different values for the parameter, in the two cases.

Relaxing our framework to a single camera setup results in a model very similar to the setup for the capture time optimization in Hasinoff et al. \cite{Hasinoff2010NoiseoptimalCF}. The only difference is in the specification of the SNR constraint. While Hasinoff et al. \cite{Hasinoff2010NoiseoptimalCF} looks to optimize the SNR of the fused image, we specify it in the form of SNR constraints on each of the captured images. We employ this greedy approach as an implementation relaxation, since our method in general allows for a variable number of images from both cameras.

\subsection{Analysis of the optimization problem}
\label{subsec:analysis}
We now describe the nature of the optimization problem on hand. In order to make the analysis tractable, the following assumptions are made:
\begin{enumerate}
    \item The ISO is assumed to be fixed.
    \item The analysis is carried out for a fixed number of primary and secondary camera images. This mirrors the nature of our optimization implementation, where the number of images for each camera is decided in the initializaion step, based on the optimization constraints. 
    \item The exposures are assumed to be continuous variables.
\end{enumerate}

First of all, the objective function (capture time) is a convex function of the exposures $t_i^j$. We make the following observations regarding the constraints. 

\begin{claim}
    Under the assumption that the radiance range of interest is a connected, continuous interval defined by its minimum and maximum values, the radiance coverage constraint is convex in the exposure times $t_i^j$.
\end{claim}

\begin{claim}
    The disparity error constraint is, in general, non-convex in the exposure times $t_i^j$. For certain scene radiance distributions, however, the constraint may reduce to be convex. 
\end{claim}

The detailed proofs for the above claims are presented in the Appendix. Based on these observations, an initialization scheme is proposed to account for non-convex cases. The initialization scheme is followed by iterative optimization to achieve local optima resulting in the final exposure sequence. 

\subsection{Our Initialization Scheme}
\label{subsec:init}

The initialization involves identifying exposure times and ISOs for the images to be captured. Note that, these values must satisfy the constraints of our optimal placement problem. There are two aspects to the initialization problem: (i) selecting the exposure times and (ii) selecting the ISOs. The proposed initialization scheme is structured so as to first select the exposure times for a given ISO configuration, and then selecting the ISO values which reduce the overall capture time.

We first look at the exposure selection scheme. For a given ISO configuration, we should choose the exposure times such that they satisfy the constraints of radiance range coverage and bounded disparity error (as mentioned in Section~\ref{subsec:criteria}). For ease of description, we modify the notation slightly from Section~\ref{subsec:criteria}. We use $t_i$ to represent the exposure time and $g_i$ to represent the gain for the corresponding image, $i \in \{0,1,...,i_{max}\}$, with even $i$ for primary camera images, and odd $i$ for secondary camera images. The indexing notation indicates that we alternate between the cameras for selecting the capture parameters, i.e., selection of parameters of an image from the primary camera is followed by the same for an image from the secondary camera. This alternating step is followed only during the initialization. While capturing the optimized sequence we capture images from the dual cameras simultaneously.

The first exposure $t_0$ is chosen such that the log radiance interval for this image starts from the lowest log radiance value of interest, $R_{min}$ (equal to the log of the minimum radiance of interest $R_{min}=log(\phi_{min})$). From Equations~\ref{eq:I_sens} and \ref{eq:Useful_range}, we arrive at the following condition:
\begin{equation}
\label{eqn:init1}
    R_{min}=e(d_l)-log(t_0)+log(g_0),
\end{equation}
where $e(\cdot)$ is the ICRF and $d_l$ is the lowest pixel value that satisfies the noise constraint for image $i$ (relating to previously introduced notation as $e(d_l) = log(x_l)$). Subsequent images are chosen so as to meet the radiance coverage and disparity error constraints. The radiance coverage constraint is met by ensuring that no gaps in radiance coverage exist between successive images. From Equations~\ref{eq:I_sens} and \ref{eq:Useful_range},
\begin{equation}
\label{eqn:init2}
\begin{split}
    & e(d_u)-log(t_{i})+log(g_{i})<e(d_l)-log(t_{i+1})+log(g_{i+1}),\\
    & \hspace{3.2cm} i\in \{0,1,...,i_{max}\},
\end{split}
\end{equation}
where $d_u$ is the highest unsaturated pixel value for image $i$ (relating to previously introduced notation as $e(d_u)=log(x_u)$). For the disparity error constraint, we adopt a strategy of geometrically diminishing errors being introduced between successive images. That is, for a given pair of images from the primary and secondary cameras, exposure values $t_i$ and $t_{i+1}$ are chosen such that the respective log radiance intervals $K_i$ and $K_{i+1}$ satisfy,
\begin{equation}
\label{eqn:init3}
    1-\int_{O}h(x)dx= \frac{\gamma_{err}}{2^{i+1}}, \hspace{0.2cm}    O = K_{i}\cap K_{i+1}, \hspace{0.2cm} i\in \{0,1,...,i_{max}\}.
\end{equation}
Essentially, this enforces the constraint that disparity error between successive images is equal to the specified value, $\frac{\gamma_{err}}{2^{i+1}}$. This choice automatically ensures that the exposures satisfy the disparity error constraint, since,
\begin{equation}
\label{eqn:init4}
\Sigma_{i=0}^{k}\frac{\gamma_{err}}{2^{i+1}}<\gamma_{err}, \hspace{0.2cm} \forall k<\infty
\end{equation}
In general, the number of images required is a variable. The value of $i_{max}$ is determined as the smallest $i$ that satisfies the following, while also satisfying all the above mentioned constraints from Equations ~\ref{eqn:init1}, \ref{eqn:init2} and \ref{eqn:init3}:
\begin{equation}
\label{eqn:init5}
    e(d_u)-log(t_{i_{max}})+log(g_{i_{max}})>R_{max},
\end{equation}
where $R_{max}$ is the highest log radiance of interest (related to the highest radiance of interest $\phi_{max}$ as $R_{max}=log(\phi_{max})$). Based on these constraints, Algorithm~\ref{alg:1} describes the exposure initialization, for fixed (given) ISOs.

\begin{algorithm}[h!]
\caption{Exposure Selection (For a given ISO configuration)}
\begin{algorithmic} 
\REQUIRE As inputs: logRadianceRange (given by the interval $[R_{min},R_{max}]$), initISOs, maxDispErr (given by $\gamma_{err}$)
\STATE $i=0$     \%comment: image number\%
\WHILE{Entire radiance range not covered (Eqn.~\ref{eqn:init5})}

\IF{i==0}
\STATE Place image at the start of  logRadianceRange (Eqn.~\ref{eqn:init1})
\ELSE
\STATE Place the next image, so as to satisfy radiance coverage (Eqn.~\ref{eqn:init2}) and disparity error (Eqn.~\ref{eqn:init3}) constraints
\ENDIF
\STATE $i\leftarrow i+1$

\ENDWHILE
\end{algorithmic}
\label{alg:1}
\end{algorithm}

A secondary benefit of this initialization scheme is that allowing higher disparity error during initial captures enables subsequent images to have lower exposures. This reduces the overall capture time.

We now look at the ISO selection regime. As discussed in Equation~\ref{eq:SNR}, the SNR is given by:
\begin{equation}
\label{eqn:init6}
\begin{split}
    SNR^j(\phi,t_{i},g_{i}) & = \frac{\phi^2 {t_{i}}^2}{\phi {t_{i}}+\sigma_{r}^2+\sigma_{q}^2{g_{i}}^2}\\
    & = \frac{x^2}{\frac{x}{g_{i}}+\frac{\sigma_{r}^2}{{g_{i}}^2}+\sigma_{q}^2}, \hspace{0.2cm} x=\frac{\phi {t_{i}}}{g_{i}}
\end{split}
\end{equation}
We can observe that (i) SNR increases with increasing pixel intensity value (which is directly related to $x$, from Equation~\ref{eq:I_sens}), and (ii) SNR decreases with increasing ISO (which is inversely proportional to the gain $g^{j}_{i}$). Hence, from (i), if for a given ISO the worst case SNR is observed for a given pixel value, all pixel values greater than this will satisfy the SNR constraint. Additionally, as ISO changes, the range of allowed pixel values (and hence the log radiance interval covered by each image) changes. Then, using the notion of $x_l$ as defined in Section~\ref{subsec:CRF} in Equation~\ref{eq:SNR} for the SNR threshold $\eta$,
\begin{equation}
\begin{split}
    \eta &= \frac{{x_l}^2}{\frac{x_l}{g_{i}}+\frac{\sigma_{r}^2}{{g_{i}}^2}+{\sigma_q}^2}, \hspace{0.2cm}\\
    x_l &=\frac{\eta}{g_i}+\sqrt{\frac{\eta^2}{{g_i}^2}+4\eta\left(\frac{{\sigma_r}^2}{{g_i}^2}+{\sigma_q}^2\right)}
\end{split}
\end{equation}

An increase in ISO (equivalent to a decrease in $g_i$) therefore results in an increase in $x_l$, while $x_u$ remains constant (since it is defined on the basis of pixel saturation). Hence, an increase in ISO results in reduced log radiance interval for a particular image. This sets up a trade-off: by increasing the ISO, while the exposure times for individual images will reduce, the reduced log radiance interval per image may lead to a requirement of capturing more image to cover the radiance range. As a result, if ISO is increased naively, while capture time for individual images reduce, overall capture time may increase. 

Algorithm \ref{alg:2} uses the above knowledge and governs the overall optimization, including ISO control. As input we provide initial ISOs for the captures. For each image, these are chosen as the lowest possible ISOs such that there exist exposures supported by the camera hardware to meet the radiance coverage (Equations~\ref{eqn:init1},
\ref{eqn:init2} and \ref{eqn:init5}), disparity error (Equation~\ref{eqn:init3}) and SNR constraints (Equation~\ref{eqn:init6}). The ISO value for each image to be captured is then increased (while keeping ISOs for other images constant), as long as the overall capture time reduces. Once the capture time ceases to reduce, the ISO is fixed, and the optimization is repeated for all images required to cover the radiance range of interest. As mentioned before, the number of images required may change with the ISOs. Hence, if more images are required by Algorithm~\ref{alg:1} on changing the ISO, the initial ISO for the new image is chosen to be the same as that of the previous image from the same camera. 

\begin{algorithm}[h!]
\caption{Overall Initialization (with ISO selection)}
\begin{algorithmic} 
\REQUIRE As inputs: logRadianceRange, initISOs
\STATE $i=0$     \%comment: image number\% 
\WHILE{All images not covered}
\WHILE{Overall capture time decreases}
\STATE Change to next available ISO for image $i$ 
\STATE Identify exposure times using Algorithm~\ref{alg:1}
\ENDWHILE
\STATE $i\leftarrow i+1$

\ENDWHILE
\end{algorithmic}
\label{alg:2}
\end{algorithm}
These initial exposures and ISOs are then fine-tuned using the Levenberg-Marquart Algorithm \cite{LMAlgo} to further reduce the overall capture time. Note that since changing ISOs may lead to a change in the number of images to be captured, the optimization treats only the exposures as variables, while keeping the ISOs constant.

\section{Pipeline for HDR and Depth}
\label{sec:Pipeline}
The proposed pipeline, as shown in Figure \ref{fig:framework}, uses the optimal exposure and ISO sequence to estimate the scene HDR and depth map. Algorithm \ref{alg:3} qualitatively describes the high level operation of the pipeline.

\subsection{Scene Radiance Distribution Estimation}
\label{subsec:radDistrEst}
To utilize the optimization framework described earlier, the radiance distribution for the scene must be estimated. We estimate the scene radiance by capturing multiple images of the scene from both the cameras. Since the baseline between the cameras is small, we need not perform registration. We assume that the fields of view of the two cameras will observe similar radiance distributions. By using ICRFs estimated a-priori for the two cameras (one-time offline estimation), we map the intensity values to radiance, thus, obtaining the scene radiance distribution. 


Note that above process of capturing multiple images from dual cameras is time consuming. 
Alternatively, we may come up with other faster approaches for radiance distribution estimation, using relevant data priors. 
Identifying suitable algorithms for the approximate scene radiance distribution estimation is left for a future work. Section~\ref{sec:disc} addresses some of these possible future directions. Since the main focus of this work is the validation of the optimization framework, 
in all further experiments, we use the accurate scene radiance distribution obtained by capturing multiple images from the stack. 

\subsection{Iterative Disparity and ICRF Estimation}
In order to obtain accurate disparity and ICRFs for the two cameras, we propose an alternating estimation setup. We initialize the ICRFs with one-time offline estimated ICRFs, from single cameras. These estimates require further improvement due to possible scale errors between the two ICRFs. In the first step of each iteration, estimated ICRFs are used to transform the images to radiance space. These images are then tone-mapped and disparity maps are computed from the tone-mapped images. In the second step, the ICRF estimate is refined using the point correspondences from the disparity map of the previous iteration. The iterations are continued until both the disparity map and the ICRFs stop improving considerably. We now describe the above steps in more detail.

\begin{figure*}
    \centering
    \includegraphics[width=\textwidth]{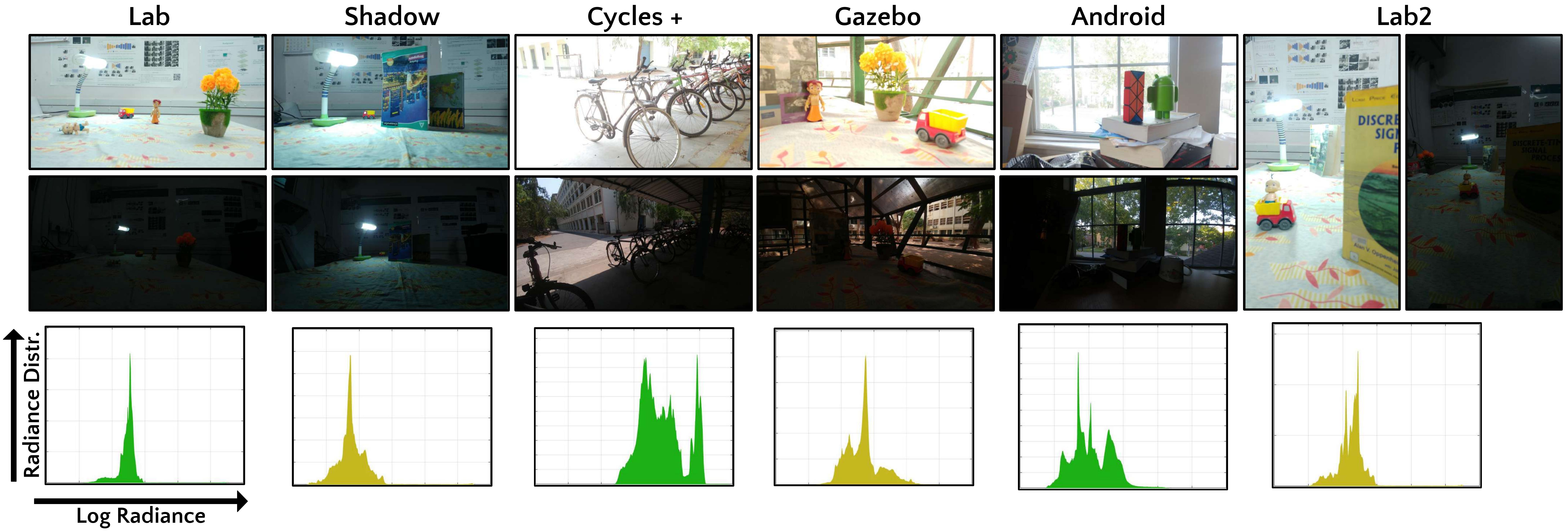}
    \caption{Scenes captured as part of our stereo HDR dataset: The top and bottom rows, respectively, show the narrow (primary) views and the wide (secondary) views. The third row shows the log radiance distribution for the captured scenes. These scenes were captured as part of our endeavor to collect a dense exposure-ISO stack of stereo HDR images using the dual camera setup of a LG G5 smartphone. We capture images for 6 scenes with an average of 142 images per scene.}
    \label{fig:scenes_used}
\end{figure*}

\noindent\textbf{ICRF Estimation:} Devebec et al. \cite{Debevec97recoveringhigh} showed that estimating the ICRF for a single camera reduces to a least squares estimation problem. 
This can now be extended to our case of dual cameras. As used previously, let $d_{i,p}^{j}$ be used to represent the image pixel values, where $i$, $p$ and $j$ refer to the image number, scene point, and the camera index respectively. From Equations~\ref{eq:I_sens} and \ref{eq:Useful_range}, we have,
\begin{equation}
\label{eqn:iterdisp1}
    R_{p}=e^j(d^j_{i,p})-log(t^j_i)+log(g^j_i)
\end{equation}
where $e^j(\cdot)$ is the ICRF for camera $j$, $R_p$ is the scene log radiance for point $p$, and $t^j_i$ and $g^j_i$ are the corresponding exposure time and gain respectively. 
We therefore define costs $C^j$ by using Equation~\ref{eqn:iterdisp1} in a weighted squared error form. Additionally, $C^j$ also includes a smoothness constraint on the ICRF to be estimated. Then the overall cost to be minimized can be given as, 

\begin{equation}
    \begin{split}
    &\hspace{2.0cm} C = C^1+C^2,\\
    C^{j} = & \Sigma_i\Sigma_{p\in S^j_i}(w(d^j_{i,p})(e^j(d^j_{i,p})-R_{p}-log(t^j_i)+log(g^j_i)))^2\\
    & +\lambda_{sm}\Sigma_{l\in \{1,...,254\}}(w(l){e^j}^{''}(l))^2,
    \end{split}
\end{equation}
 $S^j_i$ is the set of points in the $i^{th}$ image of the $j^{th}$ camera, that are valid for estimation (i.e. neither saturated nor noisy). $w(\cdot)$ refers to weighting functions, and $\lambda_{sm}$ refers to the regularization parameter. We use the triangular weighting function proposed in \cite{Debevec97recoveringhigh}.

The optimal primary and secondary camera ICRFs are therefore identified by optimizing this objective using least squares. However, in practice, it is observed that estimation of $512$ variables (for an $8$ bit image) for the two ICRFs, in addition to $R_p$ for all the scene points used for the estimation, leads to an under-determined system. We therefore apply a further relaxation step in order to make the estimation more tractable. The two ICRFs to be estimated are assumed to differ only by a constant offset factor:
\begin{equation}
    e^2(d) = e^1(d)+c,
\end{equation}
where $e^2(\cdot)$ is the ICRF for the secondary camera, $e^1(\cdot)$ is the ICRF for the primary camera, $c$ is the offset factor and $d$ is the pixel intensity value, ranging from $0$ to $255$. With this relaxation, we reduce the number of ICRF variables to be estimated to $257$.

\noindent\textbf{Disparity Estimation:} Let the $L_i^j$ be the the $i^{th}$ LDR image captured using the $j^{th}$ camera. Using the estimated ICRFs from the previous step, we transform the LDR images to log radiance space, using the ICRFs. We denote these images as the radiance-space HDR images $P_{i}^{j}$.
We then estimate the HDR image for each of the cameras, by fusing the various LDR images captured using that that camera. Lets denote these per-camera HDR images by $Q^{j}=fuse_{i=1}^{m}P_{i}^{j}$, where $fuse$ is an appropriate image fusion operator, that chooses unsaturated radiance values from the input images, wherever possible. These HDR images are then tone-mapped and the disparity map is estimated from these tone-mapped images. 

In order to improve the disparity estimation, we introduce the notion of \emph{Simulated Saturation}. Each image of the camera pair is appropriately thresholded (in radiance space) so that both images occupy the same radiance range. 
If the radiance range of $Q^j$ is $[q^j_l,q^j_h]$ this operation can be represented as follows (for $j\in \{0,1\}$), for each pixel $v$: 
\begin{equation*}
    \begin{split}
        \forall \phi \in Q^j \hspace{0.2cm}& if\hspace{0.2cm}(\phi>min_{j}(q^j_h)), \hspace{0.2cm}\phi=min_{j}(q^j_h)\\
        & if\hspace{0.2cm}(\phi<max_{j}(q^j_l)) \hspace{0.2cm}\phi=max_{j}(q^j_l).\\
    \end{split}
\end{equation*}
Essentially, pixels with radiance values outside of the common radiance range of both the cameras are synthetically saturated, by thresholding their radiances to the end-values of the common range. As a result, the tone-mapping process will be similar for both the images. However, this induced accuracy in tone-mapping is at the cost of saturating out the radiances which are not present in both the images. It is experimentally observed that the loss of radiance information is sufficiently compensated by the increased accuracy of tone-mapping. Section \ref{subsec:simSat} analyzes the benefit of using simulated saturation in our case in terms of improved disparity estimates. The performance of the ICRF estimation step and the iterative ICRF and Disparity estimation step are further addressed in Section \ref{subsec:ICRF} and \ref{subsec:ICRFandDisp}, respectively.

Note that we do not look at optimizing the proposed pipeline to work with specific disparity estimation algorithms. Instead, we aim to create a general, modular optimization framework. Thus, future improvements in disparity estimation algorithms may be easily integrated into our proposed framework. For the purpose of our experiments, we use the disparity estimation algorithm proposed by \cite{MozerovDisp}.

\begin{figure*}[t]
\begin{center}
\includegraphics[width=\textwidth]{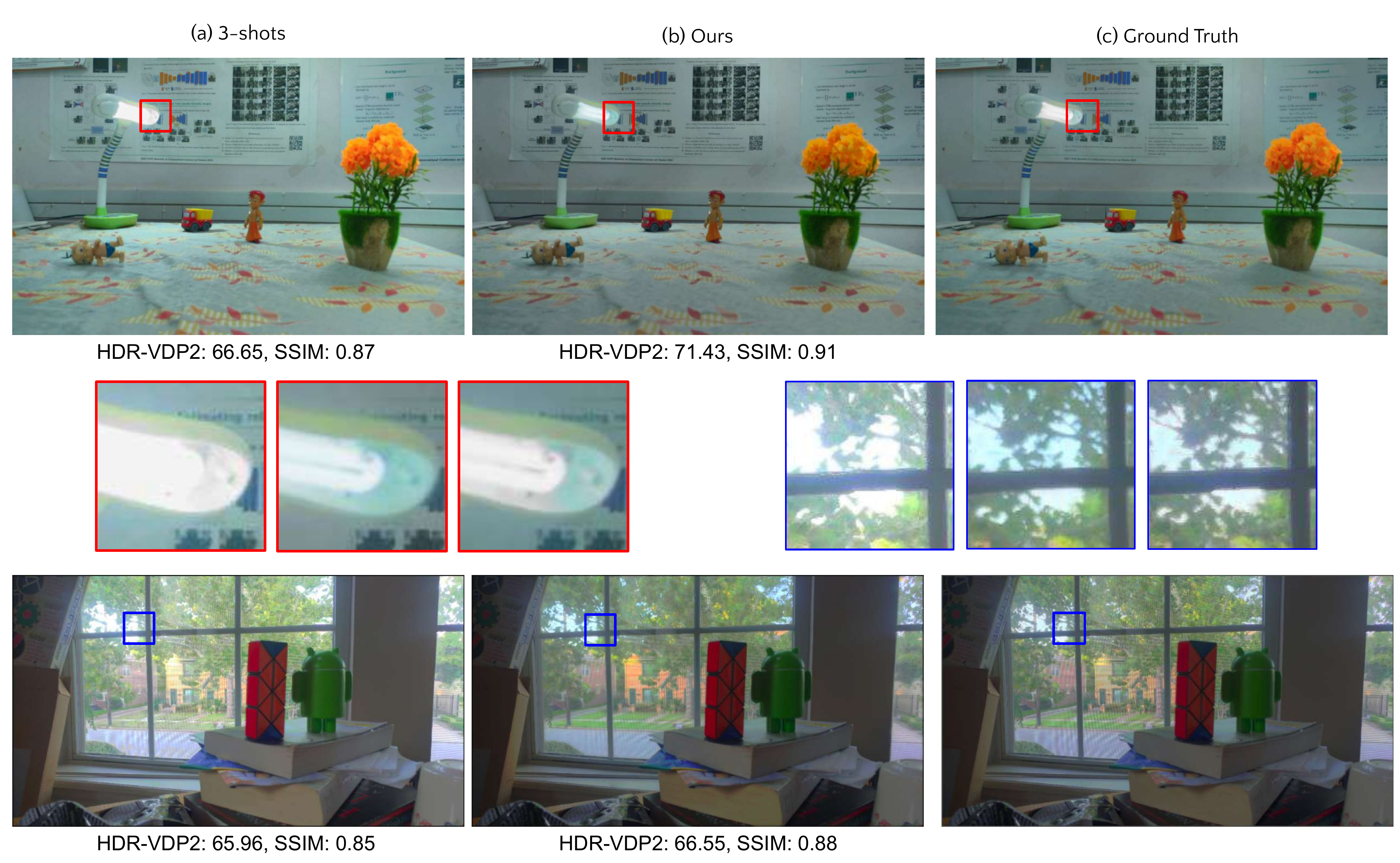}
\caption{Performance of proposed method for only HDR capture: Figure compares the performance of our StereoHDR setup in the case of HDR only reconstruction with 3-shot HDR and ground truth (GT). The 3-shot HDR is obtained from a single camera with exposure bracketing. Ground truth is obtained by using the entire exposure stack captured on the narrow view camera. Our approach from optimal input sequence is able to recover the HDR details like the details of the light bulb in the red inset, sky color in the blue inset. The metrics reported below the figures are HDR VDP-2 and SSIM respectively.}
\label{fig:hdronly}
\end{center}
\end{figure*}

\subsection{Image Fusion} 
The disparity estimates and the fine-tuned inverse Camera Response Functions can be used to fuse the radiance information into the consolidated HDR image. The disparity estimates are used to warp the secondary camera images into the primary camera view, while the ICRFs are used to transform all the captured images into the radiance space. For every pixel, the primary camera images are first checked for an unsaturated, non-noisy radiance estimate. In case this is not obtained, then secondary camera images are considered. Such a regime is followed in order to minimize stray occlusion-induced artifacts from the warped secondary view image. Additionally, in order to avoid artifacts due to radiance estimates from different images, gaussian smoothing averages are applied prior to fusion. Finally, to render the HDR image for display, a tone-mapping is applied to the fused HDR image. 
\begin{algorithm}[h!]
\caption{Stereo HDR Pipeline}
\begin{algorithmic} 
\REQUIRE \textit{As inputs: Initial ICRFs, primaryCamImgs, secCamImgs, exposures, ISOs}
\STATE Estimate initial Disparity (using initial ICRFs)
\FOR{number of iterations}
\STATE Estimate Disparity (using most recent ICRF estimates)
\STATE Estimate ICRFs
\ENDFOR
\STATE Warp secCamImgs to primary view
\STATE Convert to Radiance using ICRFs, exposures, ISOs
\STATE Image Fusion
\end{algorithmic}
\label{alg:3}
\end{algorithm}

\section{Dual camera stereo HDR dataset}
For our experiments, we require stereo camera images. However, there are few such publicly available datasets with sufficient degrees of freedom along the exposure, ISO as well as spatial (stereo) domains. We have therefore created a comprehensive stereo HDR dataset using the dual cameras from a LG G5 cell-phone. One of the cameras is a (relatively) narrow angle camera and the other a (relatively) wide angle camera. We choose the first one as the primary or reference camera, to which we map the secondary camera.  

 The dataset comprises of $6$ scenes. For each, we have obtained several exposure and ISO sequence as supported by the camera (exposures from 3.2 to 1/4000 s, ISOs from 50 to 400), resulting in an average of $142$ images per scene. This comes to approximately 30 exposures, over 4 ISO configurations, per scene. Figure \ref{fig:scenes_used} highlights this. For each scene, the narrow view at a high exposure setting and the wide view at a low exposure setting are shown, along with the scene radiance distribution. The radiance distribution is obtained using appropriate methods as suggested in Section \ref{sec:Pipeline}. The chosen scenes have radiance distributions with differing characteristics (unimodal, bimodal etc.) as well as differing dynamic ranges, allowing for versatility in testing. Please refer to the \href{https://tinyurl.com/stereoHDRsupp}{Supplementary material} for intuition on the interaction of captured images with scene radiance distributions.
 
 Before running our pipeline, the dataset images are appropriately rectified, in order to account for the differing fields of view. Additionally, effects of lens non-idealities such as radial and tangential distortions are appropriately corrected.

\section{Experiments and Results}

In this section, we perform qualitative and quantitative evaluation of our proposed framework and pipeline. Our densely captured stereo dataset is used for this purpose. 
Exposures and ISOs are optimized over the entire range of allowed values supported by the camera. The noise parameters for the cameras, specifically the read noise, quantization noise and camera gain, which are required for the optimization, were evaluated using conventional photography techniques. As mentioned previously, the disparity algorithm from \cite{MozerovDisp} was used for all our experiments. 

For quantitative comparison of recovered HDR images, we use the SSIM metric on tonemapped images, and the HDR-VDP-2 metric \cite{Mantiuk:2011:HCV:2010324.1964935} on the non-tonemapped radiance map. The HDR-VDP-2 is more relevant in terms of HDR image quality, since it is independent of the tonemapping operator used, and looks at perceptual image quality. For an exact formulation of the metric, see  \cite{Mantiuk:2011:HCV:2010324.1964935}. For disparity error, we use the percentage of scene points with inaccurate disparity as the metric. We set a threshold of $4$ pixels, and any point with a disparity deviating from the ground truth by a value larger than this is classified to have incorrect disparity.

To formalise the comparison sequences used, we draw on the notion of exposure bracketing. Here, the effective exposure between successive captured images $(i, i+1)$ is related by a constant multiplying factor. That is, $t_{exp}^{i+1}=t_{exp}^{i}\times 2^{C}$, where $t_{exp}^{i}$ is the effective exposure of the $i^{th}$ image from a particular camera, and $C$ is the exposure compensation factor.
We begin by analyzing and establishing the optimality of our proposed capture sequence selection framework. The analysis is carried out both for the only HDR recovery setup (where most existing methods perform) and for the joint disparity and HDR recovery setup (our novelty). 

\begin{table}[t]
\centering
 \begin{tabular}{ |p{2.5cm}|p{1.0cm}|p{1.1cm}|p{1.2cm}|}
 \hline
  Capture Scheme& HDR-VDP-2 & Capture Time (s) & Speedup Over GT (x1) \\
 \hline
 \hline
 3 Shot   & 61.428 & 0.300 & 1.05 \\
 \hline
 Ours & \textbf{68.550} & \textbf{0.206} & \textbf{1.42} \\
 \hline
 Full Stack (GT) & - & 0.319 & - \\
 \hline
\end{tabular}
\caption{Average comparison results for only HDR capture: Comparison is carried out with the '3 Shot' capture sequence, with respect to the single camera ground truth (full stack).}
\label{table:1}
\end{table}

\begin{figure*}[t]
    \centering
    \includegraphics[width=0.8\textwidth]{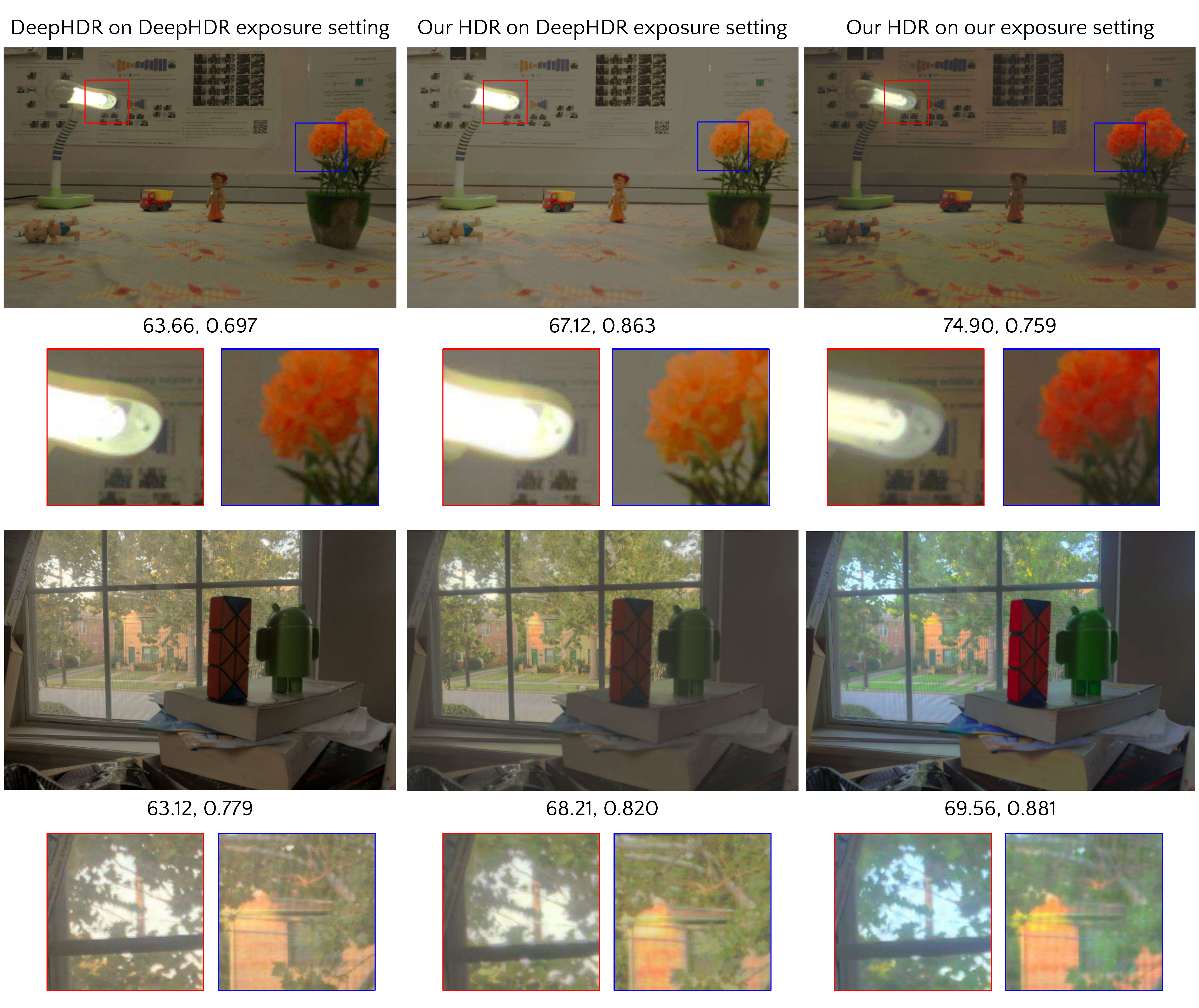}
    \caption{Comparison with DeepHDR for only HDR recovery task: First column shows DeepHDR reconstruction with their input settings, second column shows our HDR with DeepHDR input settings and the last column shows our reconstruction from our optimal input sequence. Notice that our optimal sequence recovers the details well as can been seen from the zoomed patches.}
    \label{fig:deepHDRcompar}
\end{figure*}

\subsection{Optimal HDR using dual cameras}

We first analyze the performance of our framework for the only HDR recovery task. While a single camera is sufficient for estimating the HDR image alone, using dual cameras enables the possibility of quicker capture. From the perspective of our framework, we can allow more disparity error as disparity is required only for saturated scene points in primary camera images. For the optimization scheme for the only HDR case, following notation from Equation~\ref{eqn:optISO}, we use maximum allowed disparity error of $\gamma_{err}=30\%$ and the worst case allowed SNR of $\eta=3.2$ dB .  

With the above optimization parameters, for all the dataset scenes, we obtained the optimal sequence to be consisting of $3$-$4$ images over the two cameras. We compare our results exposure bracketing scheme, which is a popular capture scheme for generating HDR image from a single camera. We consider the $3$-shot capture scheme, in which $3$ images are captured with $1$ stop exposure compensation. As a ground truth (GT), we used the HDR obtained using the full exposure stack captured with $2$ stop compensation from the primary camera. Figure \ref{fig:hdronly} shows this comparison. Our reconstructions are able to capture the dynamic range better than the $3$-shot approach, and show performance close to the ground truth. This is particularly visible from the zoomed patches. The details of the light bulb in the first scene are captured well in our reconstruction. Additionally, for the outdoor region in the second scene, our method captures the radiance accurately, whereas, the 3-shot result is saturated. This performance is further emphasized by our markedly improved SSIM and HDR-VDP-2 metrics compared to the $3$-shot approach. Our algorithm also consistently provides the best capture time over all other image capture regimes. Table. \ref{table:1} summarizes these observations. Please refer to the \href{https://tinyurl.com/stereoHDRsupp}{Supplementary material} for scene-wise qualitative and quantitative results.

\noindent\textbf{Comparison with DeepHDR:} In order to further validate our HDR fusion pipeline (Algorithm \ref{alg:3}), we compare our HDR reconstructions with a recent learning based approach, DeepHDR  \cite{LearningHDR}. We use the pre-trained DeepHDR model for our experiments. The DeepHDR model has been trained on $3$-shot captured images with $2$-stop exposure compensation. Hence, in our experiments, we use input sequences consisting of $3$ images: $2$ from the primary camera and $1$ from the secondary camera with $2$ stop exposure compensation between them. Note that although \cite{LearningHDR} uses a single camera setup for HDR reconstruction, the problem of motion artifacts, due to camera shake and dynamic scenes, is addressed using optical flow techniques. Due to the small baseline of the dual cameras, we believe the DeepHDR pre-processing stage is able to handle these disparities. 

Figure \ref{fig:deepHDRcompar} shows this comparison. In addition to the DeepHDR output, the figure highlights our HDR image from the 3-shot DeepHDR input sequence as well as our HDR from our optimal input sequence. Our pipeline enables better or comparable HDR recovery on the DeepHDR input, as can be seen from the zoomed patches and the relevant metrics. Additionally, our optimal input, from similar number of images, is able to obtain superior HDR reconstructions.  

\subsection{Jointly Optimal HDR and Depth}
\label{subsec:joinOptHDRD}
We now look at the performance of our framework for obtaining both disparity and HDR. To the best of our knowledge, there is no previous work for selecting exposure sequences for such a capture scenario. We therefore propose various image capture regimes for comparison with our optimal sequence. These comparison sequences consist of the same number of images per camera as that of the optimal sequence for each scene ($2$ images per camera for the dataset scenes). However, they differ from each other in terms of their exposure patterns. 

We propose two broad exposure patterns for comparison. For the \textit{first} one, the primary camera exposures start from the image with the lowest possible exposure time that captures the lowest radiance of interest. Each subsequent image exposure is obtained by exposure compensating the previous exposure, thereby creating a stack of images with successively decreasing exposures. Similarly, for the secondary camera, the exposures start from the image with the highest possible exposure time, that captures the highest radiance of interest, leading to a stack of images with successively increasing exposures. We show this comparison for exposure compensations of $1$, $2$ and $3$. These cases are subsequently referred to as \textit{Exp-Comp1}, \textit{Exp-Comp2} and \textit{Exp-Comp3}. For the \textit{second} sequence, the exposures are interleaved between the two cameras, by alternating between the cameras in the capture sequence. This regime is used in the comparison for exposure compensations of $2$ and $3$, subsequently referred to as \textit{Exp-Intrl2} and \textit{Exp-Intrl3}. For this case, the ground truth (GT) sequence is identified as a sequence with dense exposures along both cameras with an exposure compensation of $2$. The corresponding GT disparity is calculated between the HDR images obtained from each camera using these dense exposures. For the optimization scheme in this case, we set a maximum allowed disparity error of $\gamma_{err}=5\%$ and a worst case allowed SNR of $\eta=3.2$ dB .

\begin{figure}[t]
    \centering
    \includegraphics[trim=10 0 40 40,clip,width=0.233\textwidth]{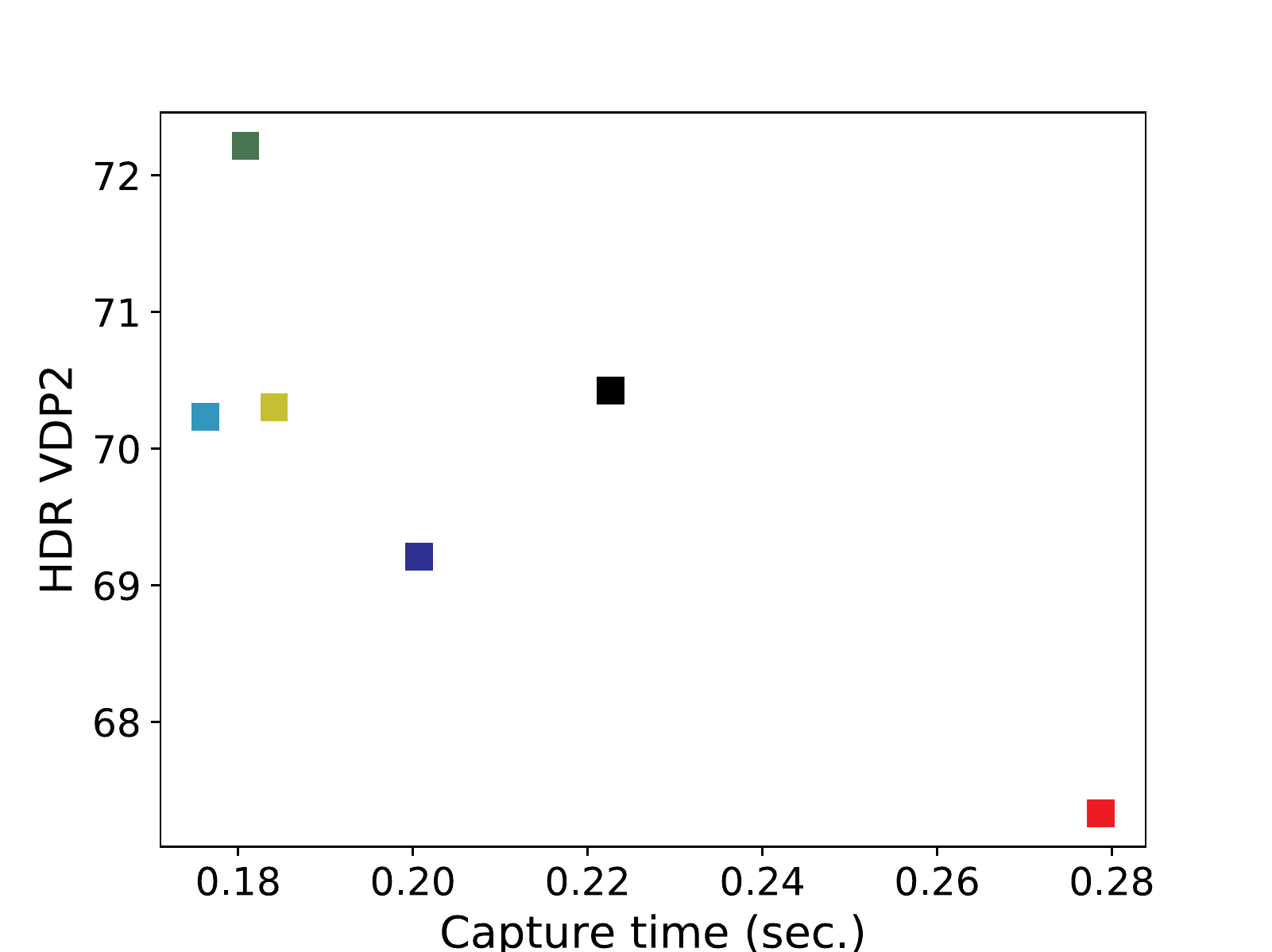}
    \includegraphics[trim=10 0 40 40,clip,width=0.233\textwidth]{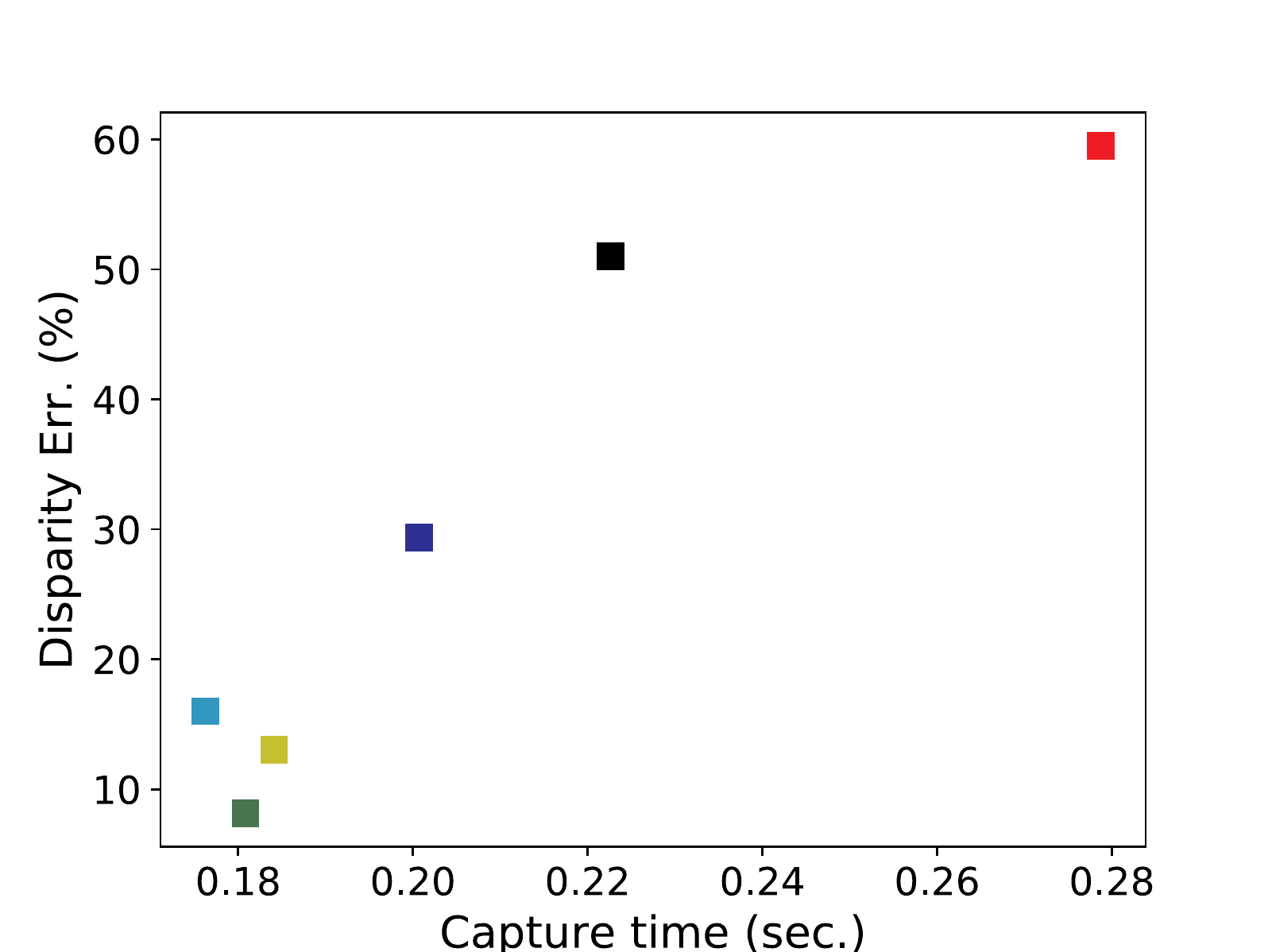}
    \vspace{0.1cm}
    \includegraphics[width=0.43\textwidth]{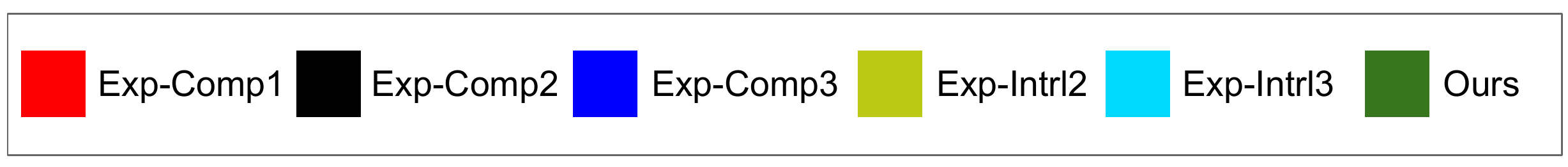}
    \caption{Comparison of our optimal sequence with other possible capture sequences ( \textit{Exp-Comp1, 2, 3} and \textit{Exp-Intrl2, 3}) for HDR and depth estimation: Plots show HDR-VDP2 \textit{vs.} Capture time and Disparity Err. \textit{vs.} Capture time. The plot shows the average results for the $6$ scenes in our dataset. 
    }
    \label{fig:scatter_plot}
\end{figure}

\begin{figure*}[t]
    \centering
    \includegraphics[width=\textwidth]{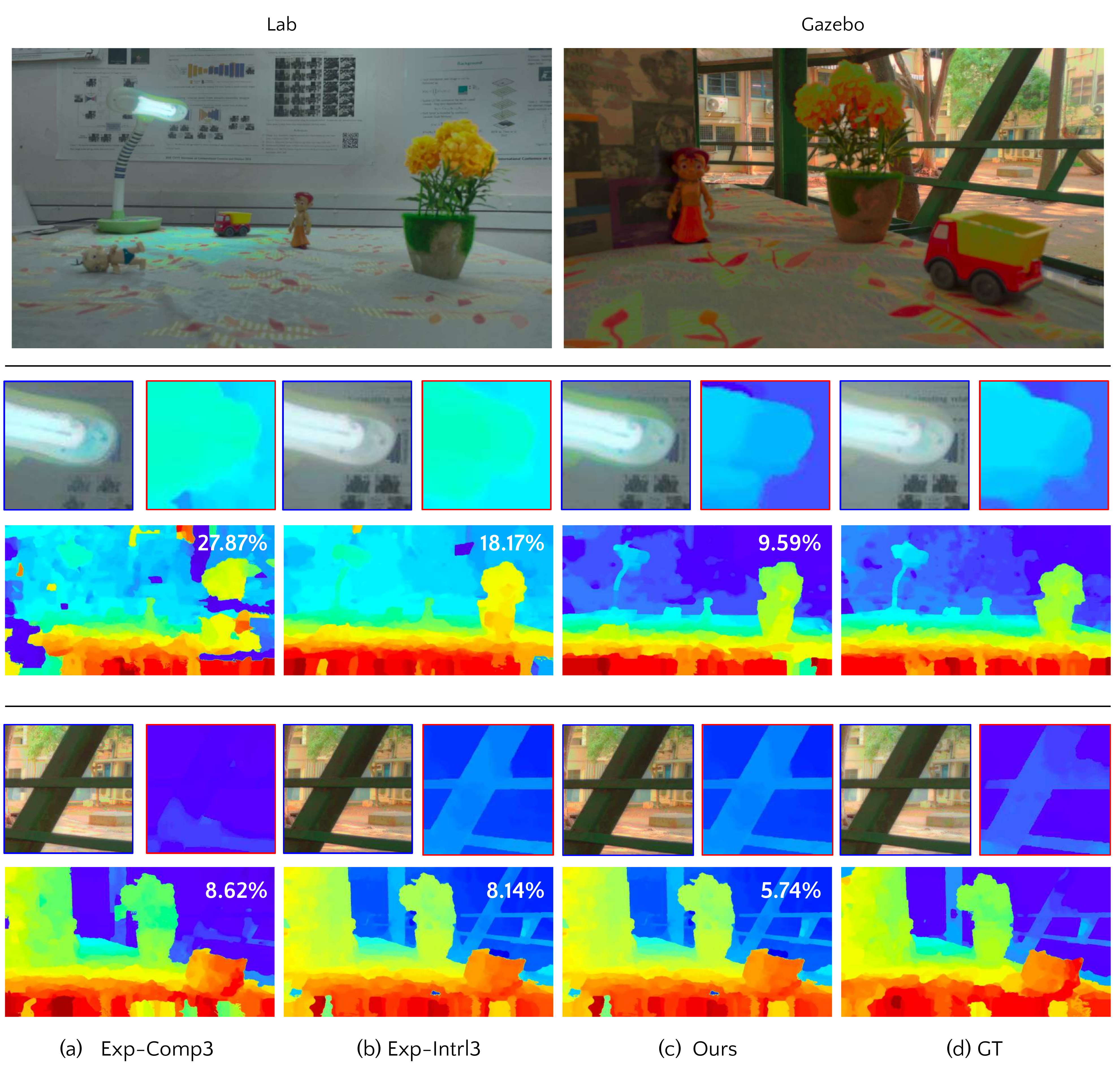}
    \caption{Comparison of HDR and depth reconstructions for various capture schemes: Top row shows HDR reconstruction of two scenes (Lab and Gazebo) for which we show the disparity maps below in the 2nd and 3rd rows. The disparities shown are: (a) from non-interleaved sequence \textit{Exp-Comp3} (2 images per camera), (b) from interleaved sequence \textit{Exp-Intrl3} (2 images per camera), (c) from our optimal exposure sequence, and (d) Ground Truth from full exposure stack. Insets show zoomed in portions of HDR and disparity for the corresponding sequence. Our optimal sequence enables better disparity estimation than others. The numbers overlaid show the percentage disparity error.}
    \label{fig:dispHDRopt}
\end{figure*}

Figure \ref{fig:scatter_plot} shows the performance of the proposed method and the comparison sequences. Specifically, we look at two scatter plots, disparity error \textit{vs.} capture time and HDR-VDP-2 \textit{vs.} capture time. Exp-Comp1 shows the worst performance in terms of capture time, HDR and disparity estimation. However, this performance quickly improves from Exp-Comp2 to Exp-Comp3 along all metrics. This is a result of the increasing radiance overlap, which enables better disparity and HDR. In comparison, the Exp-Intrl2 and Exp-Intrl3 sequences show improved performance on the disparity estimation front, as a result of generally improved overlap. However, our optimal sequence  performs better overall when compared to the other sequences. 
For a marginally slower capture time in comparison to the Exp-Intrl3 sequence, the optimal sequence shows considerable improvements in HDR quality and disparity error. For all other sequences, performance of the optimal sequence is better along all fronts. These conclusions can be visually validated from Figure \ref{fig:dispHDRopt}. Regions of high radiance, such as the lamp in Lab and the background in Gazebo, show much improved performance with respect to HDR image reproduction. Additionally, the disparity estimates show a clear improvement, especially along boundaries of rapid dynamic range change. Overall, performance very close to ground truth is observed for the optimal sequence, which is not observed in general for other comparison sequences. Comprehensive results for all scenes and exposure sequences may be found in the \href{https://tinyurl.com/stereoHDRsupp}{Supplementary material}.

\subsection{Pipeline Validation}
Here we analyze our pipeline for stereo HDR reconstruction from the optimal input sequence.  The pipeline involves two important stages, ICRF estimation and disparity estimation. Here, we evaluate these stages in further detail.

\subsubsection{ICRF Estimation}
\label{subsec:ICRF}

For the purpose of evaluation we obtain the ground truth ICRFs for both the cameras by capturing single camera exposure stacks and estimating the ICRF using the method proposed by \cite{Debevec97recoveringhigh}. To minimize the effect of  false point correspondences, we use the disparity map obtained from the HDR and disparity estimation ground truth setup (GT setup from Section~\ref{subsec:joinOptHDRD}), for this experiment. 

\begin{figure}[t]
\centering
\includegraphics[width=0.48\textwidth]{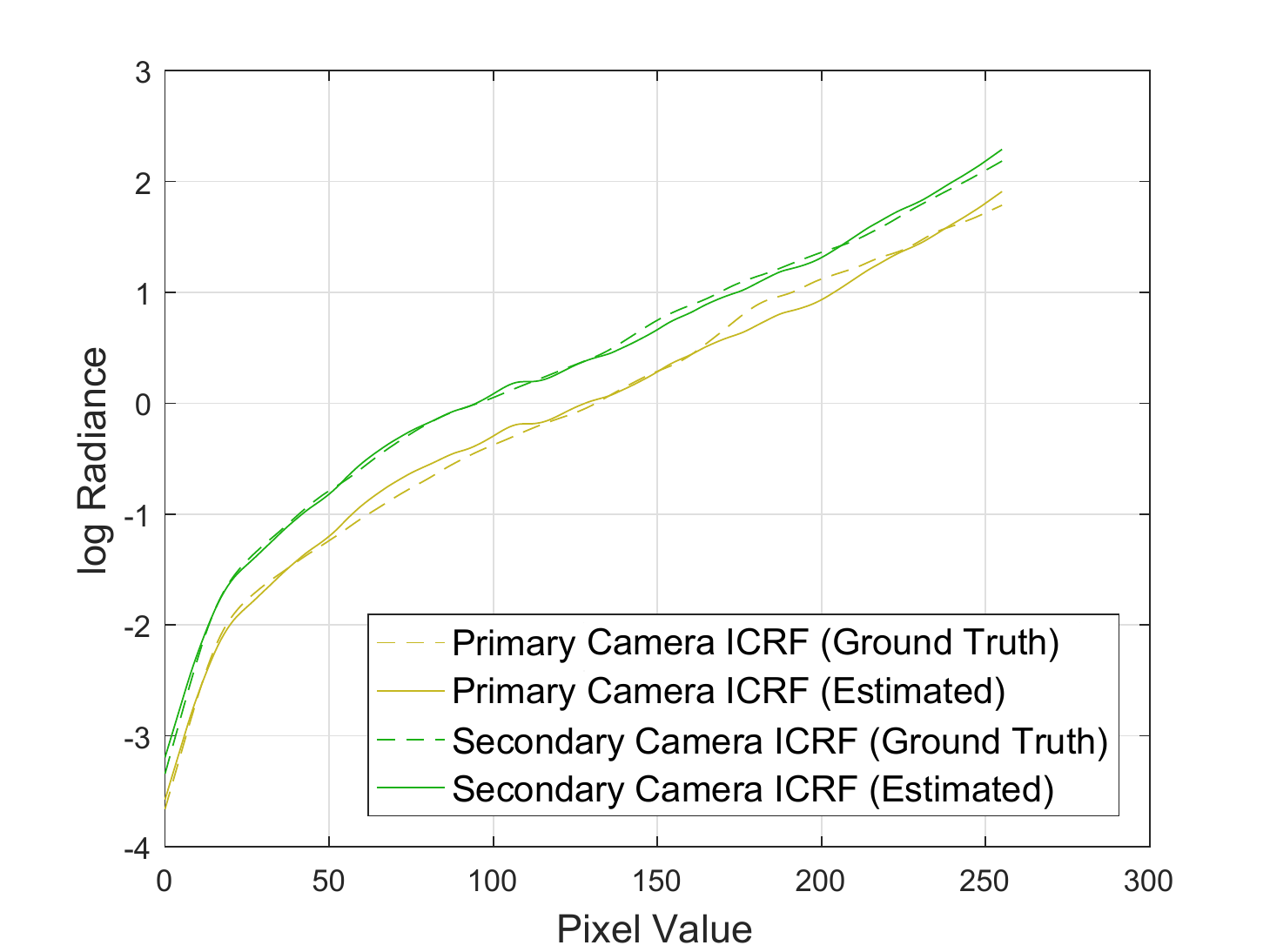}
\caption{Our estimated ICRFs are close to the desired ground truth.}
\label{fig:icrfEst}
\end{figure}

Figure \ref{fig:icrfEst} shows the relevant results. Both the primary and secondary camera ICRFs are seen to be estimated accurately, and they can be seen to closely resemble the corresponding ground truth ICRFs. Based on these results, we are able to establish that for the present setup, the relaxation of assuming identical ICRFs separated by an offset is general enough to suitably account for system characteristics.

\begin{figure}[t]
\centering
\includegraphics[width=0.48\textwidth]{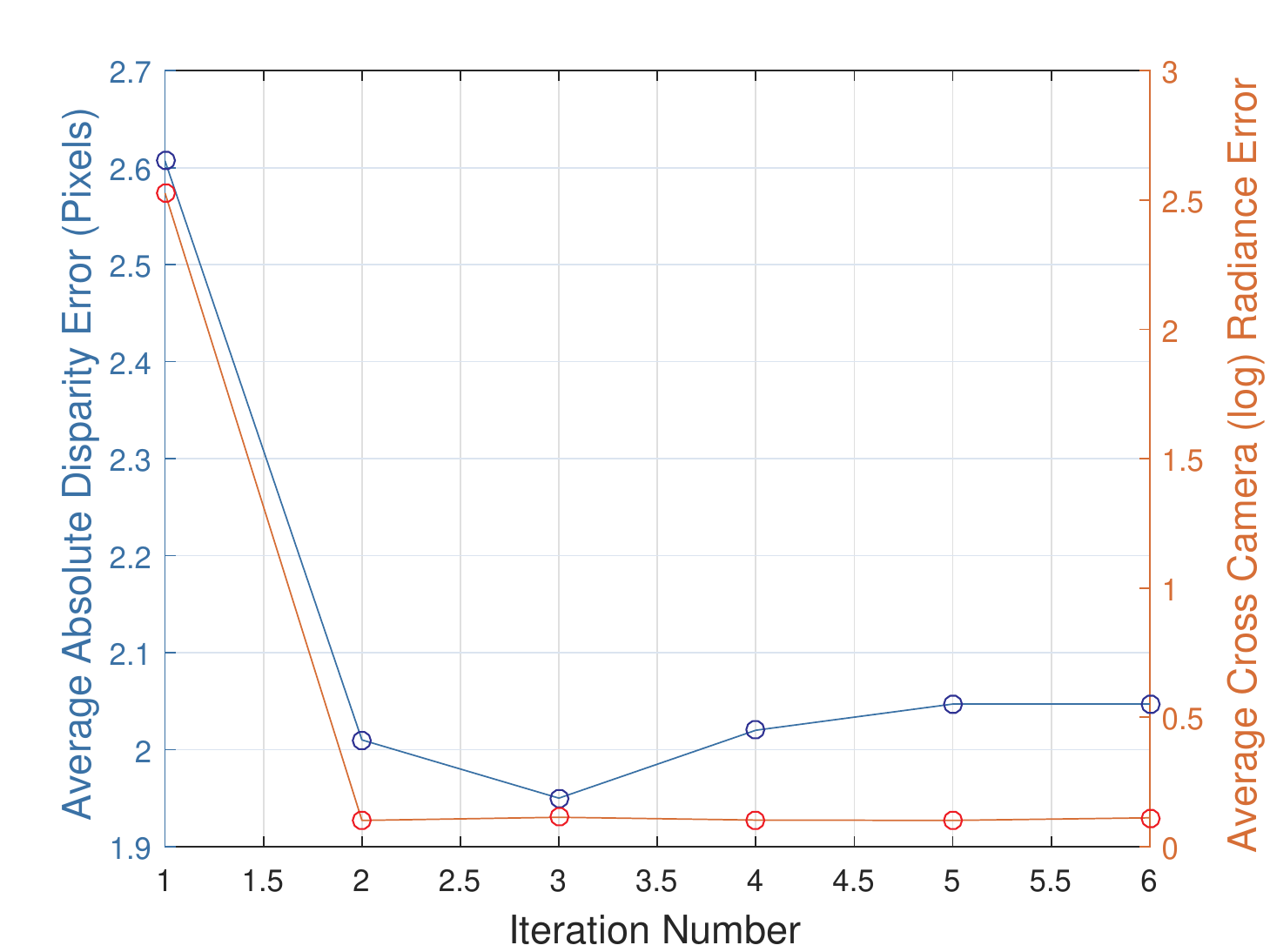}
\caption{Typical performance of the iterative ICRF and disparity estimation regime over a randomly chosen scene, in terms of mean absolute disparity error and the cross camera radiance estimation error. The benefit of the iterative scheme is visible in terms of the generally falling errors.}
\label{fig:iterative}
\end{figure}

\subsubsection{Iterative ICRF and Disparity Fine-tuning}
\label{subsec:ICRFandDisp}
We now look to validate the joint iterative ICRF and disparity estimation framework. We purposely choose corrupted initial estimated for the primary and secondary ICRFs. This is achieved by introducing a large offset in one of the ICRFs, with respect to the other. The results can be seen in Figure \ref{fig:iterative}. The error in disparity decreases over iterations. The initial error arises as a result of the corrupted ICRFs fed in as initial estimates. However, the error drops to convergence limit within $1$-$2$ iterations. The rate of convergence directly depends on the robustness of the disparity estimation algorithm in use. 
Note that the slight increase in the disparity error metric, after the initial fall, may be attributed to general stochasticity in the fine-tuning and the disparity algorithm. The overall decreasing trend is still the dominant observation.

The effectiveness of the iterations can also be seen in terms of the cross camera radiance estimation error. This metric is defined as the average error in log radiance estimated for a scene point by the two cameras. Since the initial ICRFs were corrupted, the initial error is very high. However, the error drops very rapidly to converge to a low error value. Again, the rate of convergence depends on the robustness of the disparity estimation algorithm, since the estimated disparity is the source of point correspondences for ICRF estimation. 

\begin{figure*}
\includegraphics[width=\textwidth]{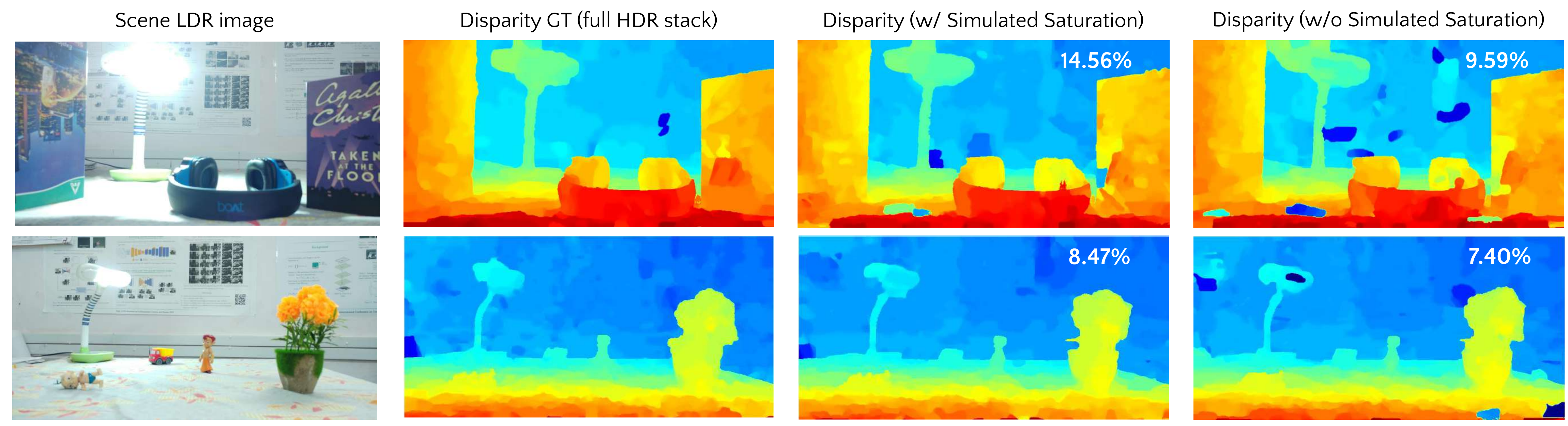}
\caption{Benefit of simulated saturation towards disparity estimation, shown for two scenes: Simulated saturation results in qualitatively and quantitatively better disparity estimates.}
\label{fig:simSat}
\end{figure*}

\subsubsection{Simulated Saturation}
\label{subsec:simSat}
Next we validate the simulated saturation step. The validation is carried out over two scenes. Using the ground truth ICRFs (for accurate analysis), we perform the radiance space conversion, followed by simulated saturation and disparity estimation. 

Figure \ref{fig:simSat} show the results. For the first scene, the lack of simulated saturation leads to spread out disparity estimation errors. For the second scene, however, the disparity estimation specifically fails for the lamp region. This drop in performance can be visually seen and is corroborated by the worse disparity error metric. Over the two scenes, therefore, simulated saturation augmented disparity estimation is able to provide much better disparity estimates.

\subsection{Applications: HDR refocusing}
\begin{figure*}[t]
\includegraphics[width=\textwidth]{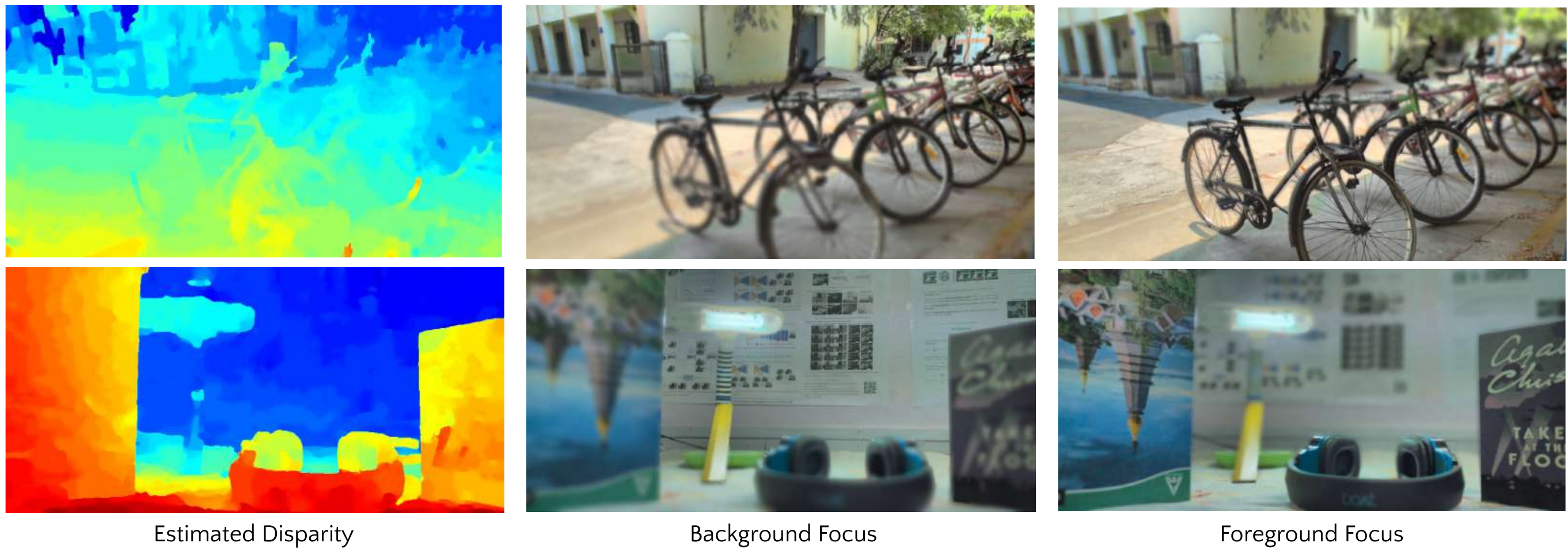}
\caption{Results of refocusing for HDR reconstructions: Desirable results may be observed from the performance on the two scenes.}
\label{fig:refoc}
\end{figure*}

We now look at HDR image refocusing as a downstream application of our optimal Stereo HDR setup. The results are shown in Figure \ref{fig:refoc} for two scenes. In both, the first image is focused on the background, while the second image is focused on the foreground. As can be seen, these results are perceptually good, since the difference between the foreground and the background can be quite easily identified and resolved. The defocus effect is consistent across the defocus region, which is desirable.

\section{Discussion and limitations}
\label{sec:disc}

The work in its current form has a few limitations, largely in the pipeline, due to major focus being on developing and analyzing the optimization framework. 
The first is the handling of occlusion. In general, this will pose a problem to the proposed method in the event of certain scene points being occluded out of the secondary camera images, while its being saturated and noisy in all the primary camera images. Such a scenario has a very low probability of occurrence in practice due to specific requirements on scene structure. However, a simple fix may be envisioned by using an occlusion aware disparity estimation algorithm in the optimal capture sequence identification step. This has not been specifically addressed here. The second limitation relates to the presence of specular highlights and isolated high radiance regions. Such regions can obviously not be represented if the hardware is unable to accommodate relevant radiances. Additionally, a very bright specular highlight may increase the overall radiance range to be captured. Since we assume that the radiances to be captured are defined by a continuous set (viable assumptions since probability distributions with zero value for finite ranges are 'almost impossible' to occur), this may lead to some wastage of resources. 

As mentioned earlier, the proposed method uses the scene radiance distribution to estimate the optimal capture sequences. For this purpose, we use dense image stacks captured from both cameras. This initial capture is not considered as part of the total capture time for the optimal sequence. 
As future work, we believe that efficient methods can be developed to estimate the distribution from fewer images. Additionally, these images may be re-used in later stages of the pipeline to bring down the capture time. Concretely, for time efficient estimation of radiance distribution, the following directions may be considered. An offline repository, containing distribution statistics of a range of candidate scenes, may be developed. For the test scene, one image each may be captured simultaneously from the two cameras, and the subset of radiance ranges estimated may be used to find the closest match from the repository. The matching may be carried out by a variation of the 'earth-movers' distance' metric. This estimated radiance distribution may then be used for HDR and depth estimation. Note that using an offline repository can also reduce compute time, since the optimal capture sequences may be pre-stored.

\section{Conclusion}
We propose a novel framework for finding the optimal exposure and ISO sequence for capturing scene HDR and depth map from a dual camera. We analyze the nature of this problem and propose an appropriate initialization and optimization scheme. We show that this optimization scheme, due to it's versatility, allows for HDR and depth as well as only HDR capture regimes.

In order to estimate the HDR and depth for the scene, we additionally propose a modular processing pipeline, which allows for the usage of task-appropriate disparity and ICRF estimation algorithms. We present the proof of concept for the same with a decent disparity estimation algorithm and a stereo-specific ICRF estimation algorithm proposed by us. The results demonstrate that the notion of an optimal capture sequence indeed holds experimental ground and our optimal exposure sequences are found to perform better than a variety of possible naive capture schemes. Additionally, our HDR-only configuration provides results which are comparable with single camera HDR techniques. We are also able to demonstrate the applicability of our framework and pipeline for the relevant downstream task of image refocusing.


%

\ifCLASSOPTIONcaptionsoff
  \newpage
\fi



%
\bibliographystyle{ieee}
\bibliography{egbib}

%
\begin{IEEEbiography}[{\includegraphics[width=1in,height=1.25in,clip,keepaspectratio]{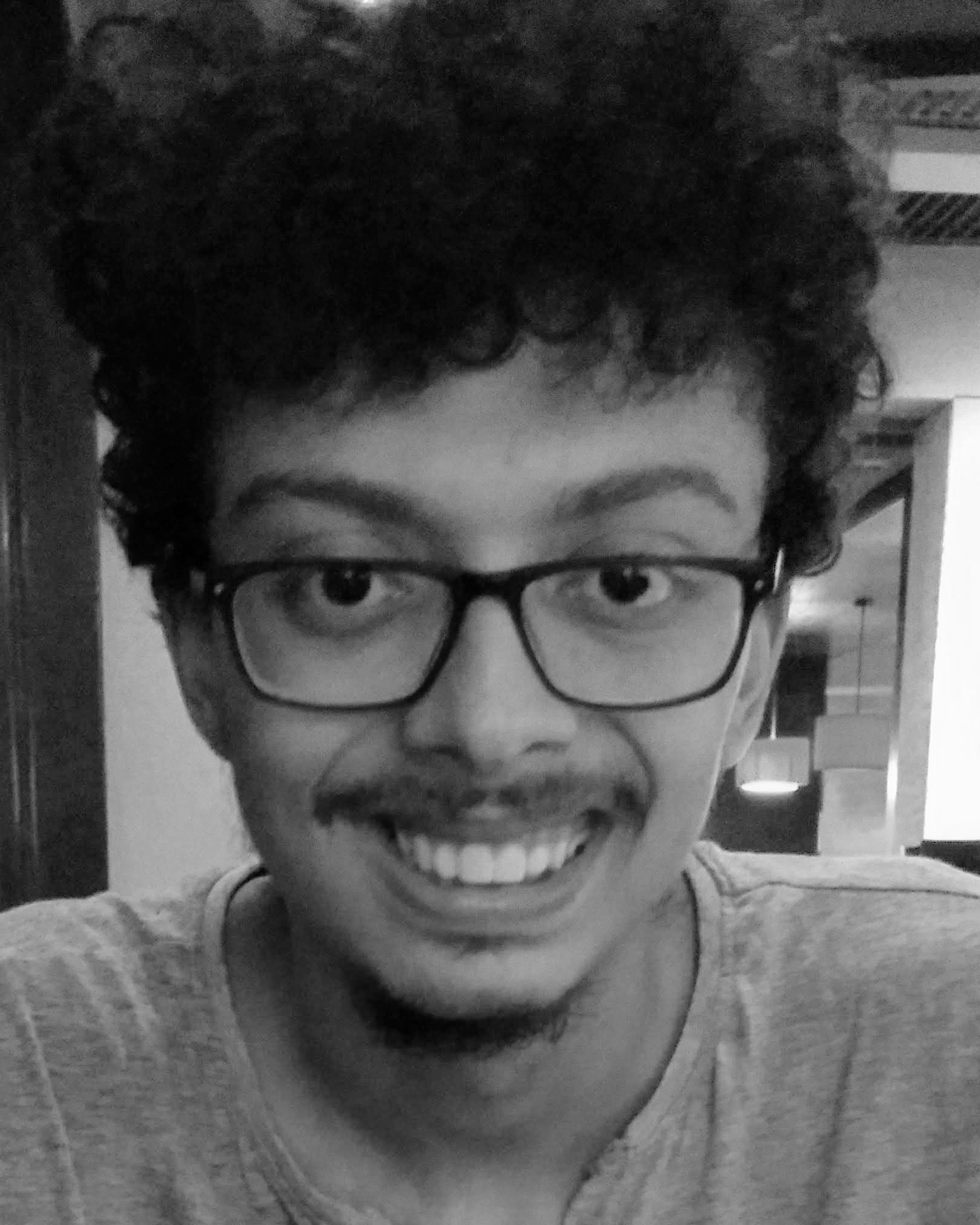}}]{Pradyumna Chari}
    received his B.Tech degree in Electrical Engineering from the Indian Institute of Technology Madras, Chennai, India. He is currently pursuing his M.S and PhD in Electrical and Computer Engineering from the University of California, Los Angeles, CA. At IIT Madras, he was associated with the Computational Imaging lab, where he worked on imaging problems for multi-camera systems. 
\end{IEEEbiography}

\begin{IEEEbiography}[{\includegraphics[width=1in,height=1.25in,clip,keepaspectratio]{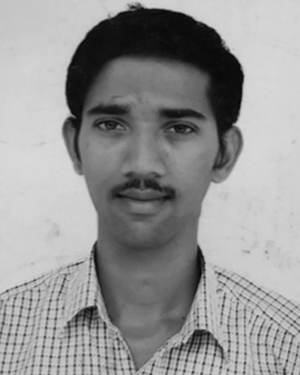}}]{Anil Kumar Vadathya}
    received the B.Tech degree in electronics and communications engineering from RGUKT, Basar, India and the M.S. degree from the Department of Electrical Engineering, IIT Madras, Chennai, India. He is currently a Research Engineer with the Scalable Health Labs, Department of Electrical Engineering, Rice University, Houston, TX, USA. He was with the Department of Electrical Engineering, IIT Madras, Chennai, India, where he worked on using learning algorithms for solving inverse problems in computational imaging. He was a recipient of Qualcomm Innovation Fellowship, India for the years 2016 and 2017.
\end{IEEEbiography}

\begin{IEEEbiography}[{\includegraphics[width=1in,height=1.25in,clip,keepaspectratio]{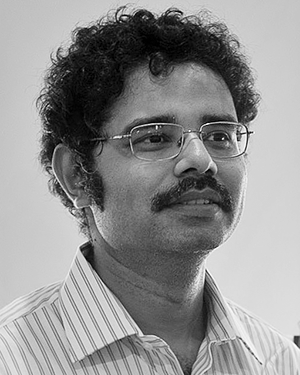}}]{Kaushik Mitra}
    received the Ph.D. degree from the Department of Electrical and Computer Engineering, University of Maryland, College Park, MD, USA. He is currently an Assistant Professor with the Department of Electrical Engineering, Indian Institute of Technology Madras, Chennai, India. Before joining IIT Madras, he was a Postdoctoral Research Associate with the Department of Electrical and Computer Engineering, Rice University, Houston, TX, USA. His research interests include computational imaging, computer vision, and machine learning. His contributions to computational imaging include proposing a theoretical framework for analysis and design of novel computational imaging systems, development of novel imaging systems, such as hybrid light field camera and assorted camera array, and using machine learning techniques, such as dictionary learning and deep learning for improving the performance of computational imaging systems.
    
    He was a Reviewer for many international journals, such as the IEEE Pattern Analysis and Machine Intelligence, the IEEE Transactions on Image Processing, IEEE Transactions on Computational Imaging, and the IEEE Journal on Selected Topics in Signal Processing. He was also a Reviewer in many international conferences, such as IEEE International Conference in Computational Photography (ICCP), SIGGRAPH, IEEE Conference on Computer Vision and Pattern Recognition (CVPR), International Conference on Computer Vision (ICCP). He was an Area Chair for the Indian Conference on Computer Vision, Graphics and Image Processing (ICVGIP), Guwahati 2016 and one of the Technical Program Committee Chairs for National Conference on Communications (NCC), IIT Madras 2017.
\end{IEEEbiography}







\newpage

\appendices
\section{Theoretical Analysis of Optimization Problem}
Here we analyze the optimization problem at hand to understand its behavior and nature. For the sake of interpretability we consider the form of the optimization problem where ISO is held constant, as in Equation 9 of the main paper. 

We first look to identify the optimization variables for the problem at hand. Our objective function is  $t_{cap}=max(\Sigma_{i=1}^mt^1_i,\Sigma_{i=1}^nt^2_i)$. Let us assume that the ICRFs for the two cameras have the functional forms $e^1:\{0,1,..,255\}\rightarrow [R^{1}_{1}, R^{1}_{2}]$ and $e^2:\{0,1,..,255\}\rightarrow [R^{2}_{1}, R^{2}_{2}]$, where $R^j_1$ and $R^j_2$ are the lower and upper log radiance limits for the image from camera $j$ (we use the indexing scheme such that cameras 1 and 2 refer to the primary and secondary cameras respectively). Then, $K^{1}_{i}=[R^{1}_{1}-log(t^1_i), R^{1}_{2}-log(t^1_i)]$ and $K^{2}_{i}=[R^{2}_{1}-log(t^2_i), R^{2}_{2}-log(t^2_i)]$ are the log radiance intervals for the two cameras, as introduced in Section~3.3 of the main paper.
The optimization problem at hand then takes the following form:
\begin{equation*}
    \begin{split}
        & Minimize \hspace{0.2cm} max(\Sigma_{i=1}^mt^1_i,\Sigma_{i=1}^nt^2_i)\\
        & Subject \hspace{0.2cm}to\\
        & \hspace{0.4cm} \cup_{j=1}^2(\cup_{i}[R^{j}_{1}-log(t^j_i), R^{j}_{2}-log(t^j_i)])\supseteq R\\
        & \hspace{0.4cm} 1-\int_{O}h(x)dx\leq \gamma_{err}, \hspace{0.2cm} \\
        & \hspace{1.4cm}O = \cap_{j=1}^2(\cup_{j=1}^m[R^{j}_{1}-log(t^j_i), R^{j}_{2}-log(t^j_i)]).\\
    \end{split}
\end{equation*}
The optimization variables are hence observed to be the exposure times $t^j_i,\hspace{0.2cm} j\in \{1,2 \},\hspace{0.2cm} i\in \{1,2,...,m_j \} $. 
Our specific assumptions for this analysis are as mentioned in Section 3.5 of the main paper. 
Our objective from this analysis is to characterize the nature of the optimization problem, in terms of convexity.
In order for the problem to be convex, (a) the objective must be a convex function and (b) the feasible space must define a convex set. 
\subsubsection{Nature of the Objective Function}
The objective function is given by $F_o=max(\Sigma_{i=1}^mt^1_i,\Sigma_{i=1}^nt^2_i)$. Since the objective consists of a 'maximum' operation on two hyperplanes, the resulting objective function is a convex function. 
\subsubsection{Nature of the Constraints}
We now analyze each of the two constraints individually for their convexity properties.\\
\\
\textbf{Claim 1- Radiance Coverage Constraint}:
Consider the radiance coverage constraint, that is, each radiance must be captured by at least one camera, without saturation. 
Let $i$ and $i+1$ represent two consecutive captures, in terms of radiance coverage. This essentially refers to two images such that the second image is the image with higher exposure that captures scene radiance values larger than those captured by the first image (closest to the first image, on the higher radiance side). Note that these two images may be from the same or different cameras. This fact does not affect the proof.


Let $t_i$ and $t_{i+1}$ be the exposure times for these two images, and let $j_i$ and $j_{i+1}$ represent the respective camera indices. Note the slight change in notation; notation in use here is camera agnostic, since the proof does not depend on the camera index.
Then, the radiance coverage constraint for this pair of exposures reduces to:
\begin{equation*}
    R^{j_{i+1}}_{1}-log(t_{i+1})\leq R^{j_i}_{2}-log(t_i),
\end{equation*}
This essentially enforces the fact that the lowest radiance captured by the second image must be less than the highest radiance captured by the first image. We now introduce the substitutions $R^{j_{i+1}}_{1}=log(T_1)$ and $R^{j_i}_{2}=log(T_2)$, the constraint becomes:
\begin{equation*}
    \begin{split}
        & log(T_1)-log(t_{i+1})\leq log(T_2)-log(t_{i})\\
        \implies & t_{i+1} \geq \frac{T_1}{T_2}t_{i}
    \end{split}
\end{equation*}

\noindent The coverage constraint for the pair of images considered is therefore found to define a half-space. Hence, for these two images, the constraint is convex. By applying a similar analysis for all pair of consecutive images, the constraint will be defined as a set of half-spaces. Since an intersection of convex sets is also convex, the overall disparity error constraint defines a convex set.\\
\\
\textbf{Claim 2- Disparity Error Constraint}:
Here, we look to analyze the disparity error constraint. We show that for some special case such as uniform radiance distribution the constraint is convex. However, in general, the constraint is non-convex. First, we look at a special case of uniform radiance distribution. 

\underline{Special case of convexity for uniform radiance distribution:}  We continue from the optimization problem defined previously. For simplicity, we change our notation of indexing exposures as follows:
\begin{equation*}
    \begin{split}
        t^1_i \rightarrow t_{2i-1},\hspace{0.2cm}
        t^2_i \rightarrow t_{2i},
    \end{split}
\end{equation*}
where odd-indexed exposure times refer to primary camera images and even-indexed exposure time refer to secondary camera images.
Let the radiance distribution for the scene be represented by the function $h(.)$. Under these circumstances, the disparity error constraint may be rewritten as:
\begin{equation*}
    \Sigma_{i=1}^{n-1} \int_{R^{mod(i,2)+1}_{1}-log(t_{i+1})}^{R^{mod(i+1,2)+1}_{2}-log(t_i)} h(x)dx \geq 1-\gamma_{err}
\end{equation*}
Here, $n$ is the total number of images from the two cameras, and $mod(.,.)$ refers to the modulus function (modulus function is brought into use for camera indexing purposes). Essentially, the disparity error is accumulated over the log radiance interval lower bounded by the lowest radiance captured by the second image, and upper bounded by the highest radiance captured by the first image. We look at the case when the scene radiance distribution $h(.)$ is a uniform distribution. In such a case, the constraint evolves as follows (note that $k$ is an appropriate constant to ensure that $h(.)$ satisfies the constraint for a probability distribution):
\begin{equation*}
    \begin{split}
        & \Sigma_{i=1}^{n-1} k(R^{mod(i+1,2)+1}_{2}-log(t_i)-R^{mod(i,2)+1}_{1}\\
        &\hspace{3.9cm}+log(t_{i+1})) \geq 1-\gamma_{err}\\
        \implies & log(\frac{t_n}{t_1}) \geq \frac{1-\gamma_{err}}{k}-k'\\
        & \hspace{1.45cm}(k'=\Sigma_{i=1}^{n-1}R^{mod(i+1,2)+1}_{2}-R^{mod(i,2)+1}_{1})\\
        \implies & t_n \geq \epsilon t_1 \hspace{0.7cm} (\epsilon=exp(\frac{1-\gamma_{err}}{k}-k'))
    \end{split}
\end{equation*}
This form for the expression satisfies the convexity constraint, since it defines a halfspace. We see that the problem at hand is convex in the case of a uniform distribution for the radiance.\\

\begin{figure}
\centering
\includegraphics[width=7cm]{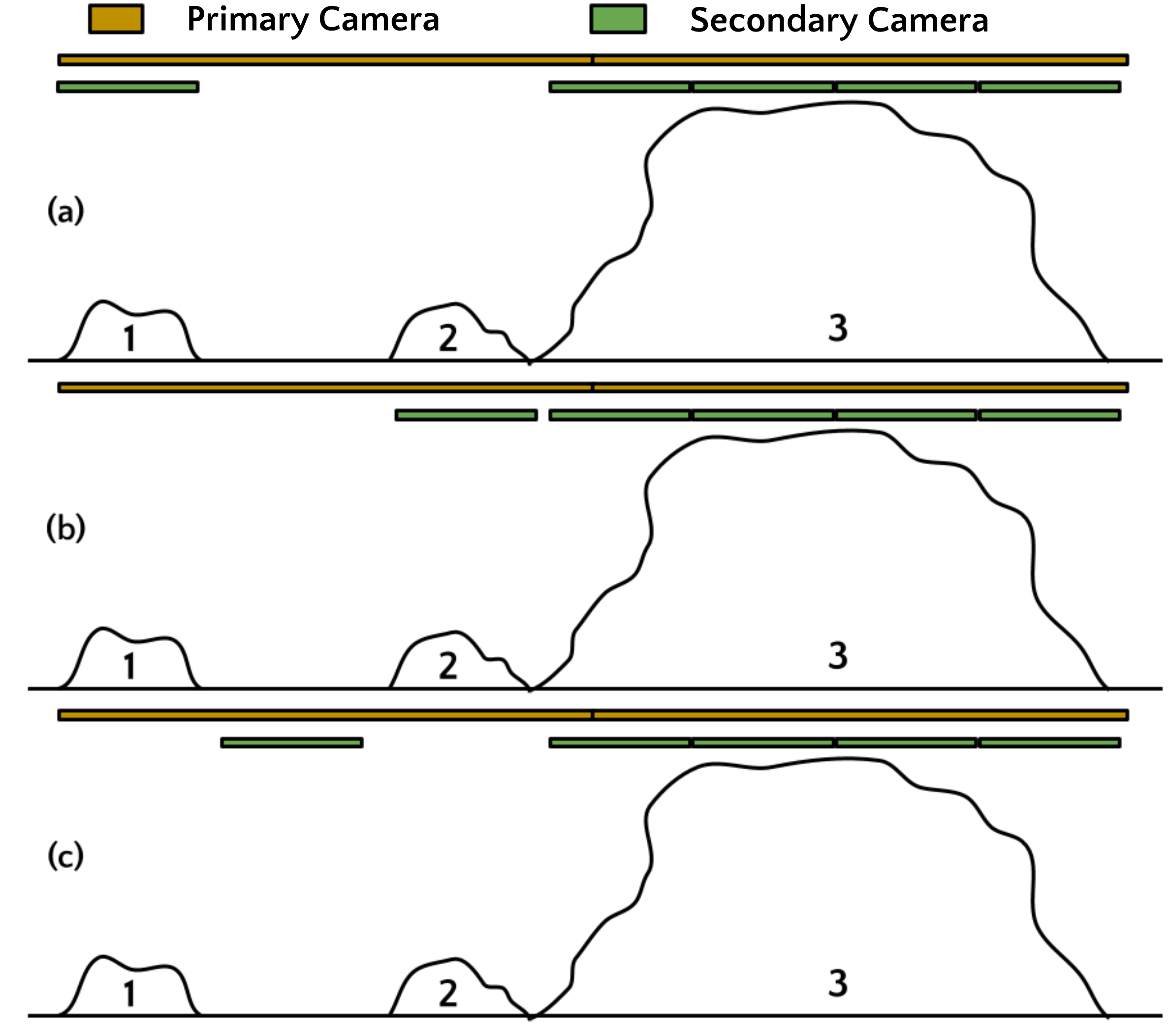}
\caption{This figure describes the exposure sequences under consideration for the proof. (a) First exposure sequence, that satisfies the constraints. (b) Second exposure sequence, that satisfies the constraints. (c) Third exposure sequence, which is a convex combination of the first two exposure sequences, that does not satisfy the constraints.}
\label{fig:seqs}
\end{figure}

\underline{General case of non-convexity:} Consider the radiance distribution as described in Figure \ref{fig:seqs}.
The scene radiance distribution consists of three regions: regions 1 and 2 have fraction of radiance equal to $\gamma_{err}$ each ($\gamma_{err}$ is the allowed error in disparity, as described in the opitmization problem). Region 3 therefore has a fraction of radiance equal to $1-2\gamma_{err}$. Additionally, the radiance ranges of the primary and secondary cameras are as shown in the figure.

Based on the above definitions, the sequences considered in (a) and (b) both satisfy the optimization constraints. In these two, all images except the first image from the secondary camera are the same. As a result, (c) represents the exposures obtained by a convex combination of the exposures from (a) and (b). Hence, (c) does not meet the disparity error requirements. For such a setting, the disparity error constraint in non-convex. Therefore, in general, the disparity error constraint may be non-convex, depending on the radiance distribution and camera radiance ranges.

\end{document}